\documentclass[aps,prd,letterpaper]{revtex4}

\usepackage{latexsym}
\usepackage{epsfig}
\usepackage{scrextend}
\usepackage[french]{babel}
\usepackage{url}
\usepackage{hyperref}
\usepackage{latexsym}
\usepackage{epsfig}
\usepackage{scrextend}
\usepackage{lipsum}

\usepackage{natbib}

\usepackage{tikz}
\usetikzlibrary{positioning,arrows,decorations.pathmorphing,arrows.meta}

\usepackage{xcolor}
\usepackage{amssymb}
\usepackage{amsmath}
\usepackage{graphicx}

 \definecolor{blue}{RGB}{7,80,201}
 \definecolor{red}{RGB}{200,20,1}

\def\be{\begin{equation}}
\def\ee{\end{equation}}
\def\bea{\begin{eqnarray}}
\def\eea{\end{eqnarray}}
\def\ba{\begin{array}} 
\def\ea{\end{array}}
\def\bc{\begin{center}}
\def\ec{\end{center}}
\def\u{\underline}
\def\ghost#1{}

\def\simge{\mathrel{
   \rlap{\raise 0.511ex \hbox{$>$}}{\lower 0.511ex \hbox{$\sim$}}}}
\def\simle{\mathrel{
   \rlap{\raise 0.511ex \hbox{$<$}}{\lower 0.511ex \hbox{$\sim$}}}}

\def\dis{\displaystyle}

\newcommand{\footremember}[2]{%
    \footnote{#2}
    \newcounter{#1}
    \setcounter{#1}{\value{footnote}}%
}

\begin{document}

\title{{\ }  \vspace{5mm}
 \hspace*{-2mm}\framebox [15cm]{\rule[-.85cm]{0cm}{2.2cm} 
$ \ba{c} \hbox{A NEW DUAL SYSTEM \,FOR THE FUNDAMENTAL UNITS}  \\ [3mm]
 \hbox{\hbox{\em  including and going beyond the newly revised} \,SI}
 \ea $
}}

\author{{\sc P}{ierre} {\sc Fayet}
\vspace{5mm} \\ \small }

\affiliation{Laboratoire de physique de l'\'Ecole normale sup\'erieure\vspace{.5mm}\\ \hspace*{4mm}24 rue Lhomond, 75231 Paris cedex 05, France 
\hspace{-1mm}\footremember{a}{\vspace*{0mm}   \em LPENS, \'Ecole normale sup\'erieure, Universit\' e PSL; CNRS UMR  8023;  
\vspace{.8mm}Sorbonne Universit\'e;    UPD-USPC}
\vspace{1.5mm}\\
\hbox{and \,Centre de physique th\'eorique, \'Ecole polytechnique, 91128 Palaiseau cedex, France}\
\vspace{3mm}}


\begin{abstract}

\textwidth 10cm

\vspace{0mm}

\bc
{\bf Abstract}
\ec

{\vspace{-2.5mm}

\hspace*{2mm}

We propose a new system for the fundamental units, 
\vspace{.2mm}
which includes and goes beyond the present redefinition of the SI, by choosing also $c=\hbar=1$.
\vspace{.1mm}
By fixing  $\,c=c_\circ $\,m/s = 1, $\,\hbar=\hbar_\circ $\,J\,s = 1 and $\, \u{\mu_\circ }=\mu_\circ $ N/A$^2$ = 1, 
\vspace{-.6mm}
it allows us to define the metre, the joule, and the ampere as equal to (1/299\;792\;458)~s, $\,(1/\hbar_\circ  = .948\,... \times 10^{34})\ {\rm s}^{-1}$ and 
\vspace{.2mm} 
$\,\sqrt{\mu_\circ \, \rm N}\,= \sqrt{\mu_\circ c_\circ / \hbar_\circ }\ \,{\rm s}^{-1}= 1.890\,...\times 10^{18}\ {\rm s}^{-1}$. 
It presents at the same time the advantages 
\vspace{.2mm}
and elegance of a system with $\,\hbar=c=\u{\mu_\circ}=\u{\epsilon_\circ }=k=N_A=1\,$, where the vacuum magnetic permeability, electric permittivity, and impedance are all equal to 1.

\vspace{.8mm}

All units are rescaled from the natural ones
\vspace{.3mm}
and proportional to the s, s$^{-1}$, s$^{-2}$, ... or just 1, as for the coulomb, ohm and weber, now dimensionless.
\vspace{-.3mm}
The coulomb is equal to\,$\sqrt{\mu_\circ c_\circ / \hbar_\circ}= 1.890... \times 10^{18}$, 
and 
\vspace{-.1mm}
the elementary charge to $\, e=1.602\,...\times 10^{-19}\, {\rm C} = 
\sqrt{4\pi\alpha}=.3028\,...\,$. 
\vspace{.3mm}
\,The ohm is equal to  $1/\mu_\circ c_\circ$ so that the impedance of the vacuum is
$\,Z_\circ\! =
376.\hspace{.2mm}730\,... \,  \Omega =1\,$.
\vspace{-.2mm}
The volt is 
$\, 1/ \sqrt{{\mu_\circ c_\circ \hbar_\circ}}\ \,{\rm s}^{-1} \!= $ $5.017\,...  \times 10^{15}\ {\rm s}^{-1}$,
and  the tesla 
$\,c_\circ $V/m  = $ \sqrt{{c_\circ^3}/{\mu_\circ\hbar_\circ}}\ \, {\rm s}^{-2} = \,4.509\,... \times  10^{32}\ {\rm s}^{-2}$. 
\vspace{2mm}

The weber is 
$ 1/ \sqrt{{\mu_\circ c_\circ \hbar_\circ}}= \,5.017\,...  \times 10^{15}$. 
\vspace{.3mm}
\,The flux quantum is $\,\Phi_\circ\! =h/2e= 
2.0678\,...\, \times 10^{-15}$~Wb 
=
$\pi/e=  10.374\,...$, with
\vspace{.3mm}
$K_J\! =483\,597.\,... $\,GHz/V$\,= e/\pi  = $ .096\hspace{.3mm}39\,..., and 
$R_K\!=25\,812.\; ...\, \Omega =1/2\alpha\simeq $ $ 68.\hspace{.3mm}518\,$.
One can also fix $e=1.602\;176\;634 \times 10^{-19}$ C, 
\vspace{.2mm}
at the price of adjusting the coulomb and all electrical units
with $\,\mu_\circ=4\pi\times 10^{-7}\,\eta^2\,$ where $\eta^2, \,\propto \alpha\,$,  is very close to 1.

\vspace{5mm}

\bc
{\bf R\' esum\' e}
\ec
\vspace{1mm}

Nous proposons un nouveau syst\`eme pour les unit\' es fondamentales, qui inclut
la red\'efinition en cours du SI et va au-del\`a, en choisissant aussi $c=\hbar=1$.
En fixant $c=c_\circ $\,m/s = 1, $\hbar=\hbar_\circ \rm \,J\,s $ = 1 et $\, \u{\mu_\circ }=\mu_\circ $ N/A$^2$ = 1, 
\vspace{-.5mm}
il permet de d\' efinir le m\`etre, le joule 
\vspace{-.1mm} 
et l'amp\`ere comme \' egaux \`a  (1/299\;792\;458)~s, $\,(1/\hbar_\circ  = 0,948\,... \times 10^{34})\ {\rm s}^{-1}$ et
$\sqrt{\mu_\circ \, \rm N}\,= \sqrt{\mu_\circ c_\circ / \hbar_\circ }\ \,{\rm s}^{-1}=$ $ 1,890\,...\times 10^{18}\ {\rm s}^{-1}$. 
\  Il pr\' esente en m\^eme temps les avantages et l'\' el\' egance d'un syst\`eme o\`u
 $\,\hbar=c=\u{\mu_\circ}=\u{\epsilon_\circ }=k=N_A=1\,$, o\`u la perm\'eabilit\' e magn\'etique, la permittivit\' e \'electrique et l'imp\' edance du vide sont \'egales \`a 1.
\vspace{1mm}

Toutes les unit\' es sont red\' efinies \`a partir des unit\' es naturelles et proportionnelles \`a la 
seconde, s$^{-1}$, s$^{-2}$, ... \ ou \`a 1,  comme pour le coulomb, l'ohm et le weber, sans dimensions.
\vspace{.2mm}
Le coulomb est \' egal \`a  $\sqrt{\mu_\circ c_\circ / \hbar_\circ}= 1,890\,... \times 10^{18}$, 
\vspace{.2mm}
et la charge \' el\' ementaire \`a 
$\, e=1,602\,...\times 10^{-19}\ {\rm C} = 
\sqrt{4\pi\alpha}\,=\,0,3028...\,$.  \linebreak
L'ohm est \' egal \`a  $1/\mu_\circ c_\circ$, et l'imp\' edance du vide \`a
$Z_\circ\! =
376,\hspace{-.2mm}730\,... \  \Omega =1$.
\vspace{-.2mm}
Le volt est 
$1/ \sqrt{{\mu_\circ c_\circ \hbar_\circ}}\ \,{\rm s}^{-1}$ $= 5,017\,...  \times 10^{15}\ {\rm s}^{-1}$,
et le tesla
$\,c_\circ $V/m  = $ \sqrt{{c_\circ^3}/{\mu_\circ\hbar_\circ}}\ \, {\rm s}^{-2} = $ $\,4,509\,... \times  10^{32}\ {\rm s}^{-2}$. 

\vspace{1.5mm}

Le weber est 
$ 1/ \sqrt{{\mu_\circ c_\circ \hbar_\circ}}= 5,017...  \times 10^{15}$, et le 
\vspace{.1mm}
quantum de flux $\,\Phi_\circ\! =h/2e= 
2,0678\,... \times 10^{-15}$~Wb =
$\pi/e =  10,374\,...$\,, avec
\vspace{.1mm}
$K_J\!=483\,597.\;... $ GHz/V $ = e/\pi =$\, 0,096\hspace{.3mm}39\,...\,, et
$R_K\!=25\,812.\;...\;\Omega =1/2\alpha 
\simeq 68,518\,$.
On peut aussi fixer $e=1,602\;176\;634\times 10^{-19}$ C
\vspace{.1mm}
\`a condition d'ajuster le coulomb et toutes les unit\' es \' electriques avec $\mu_\circ=4\pi \times 10^{-7}\, \eta^2$ o\`u $\eta^2, \,\propto \alpha$,  est tr\`es proche de 1.

\ \vspace{-.5mm} \\
\hspace*{2mm}
\\
\hbox{\hspace{110mm}LPTENS/18/19}}

\end{abstract}

\maketitle

\vspace{-9mm}

\section{Presentation of the new system} 

\label{sec:intro}

\vspace{.5mm}

Measuring geometrical or physical quantities, such as intervals of time, space distances, masses, energies, electric charges and currents, etc., requires the definition of appropriate units.
It is desirable that these units be the same everywhere, and do not change with time. 
To realize this they must be defined as much as possible in a universal way, from invariant physical objects or phenomena. This is how the metre was initially defined as ``la dix-millioni\`eme partie du quart du m\' eridien terrestre''
 \cite{metre}.  The unit of weight followed, 
 as the weight of a dm$^3$ of water at the temperature of melting ice \cite{gramme}.
This was at the origin of the metric system in France at the end of the XVIII\hspace{.2mm}th century, and later of the ``Convention du m\`etre" in 1875 \cite{convention},  leading to the International System of Units (SI) \cite{SI,DQ}.

\vspace{2.5mm}
As of today, the second is defined from the period of a specific transition of the caesium-133 atom~\cite{sec}. The metre is derived from the second by fixing the value of the speed of light in vacuum, $c$, to be exactly\,  299\;792\;458 m/s\, \cite{c}. \,This is made possible thanks to the theory of relativity, verified to a very high degree of precision, according to which time and space are related entities of similar nature, the speed of light being the same, always and everywhere, independently of the reference frame in which it is measured.  

\vspace{2.5mm}
But the unit of mass, the kilogram, still remains defined from a physical artefact, as the mass of the international prototype of the kilogram ({\boldmath $\cal K$}), a cylinder of platinum-iridium alloy stored at the Inter\-national Bureau of Weights and Measures (BIPM) in the ``Pavillon de Breteuil'' at S\`evres, France \cite{kilo}.  Its mass remains constant and equal to 1\,kg by definition and according to an international agreement, even if the corresponding quantity of matter cannot remain exactly constant due to surface effects and does inevitably vary very slightly over the years \cite{ipk}.
This is not a satisfactory situation, and it would be desirable to get a universal definition of the kilogram, based on reproducible universal phenomena rather than on a single material object.

\vspace{2.5mm}
This is the main purpose of the present redefinition of the kilogram, that will soon be derived from the second, or the metre, using quantum mechanics, by  fixing the value of the Planck constant  $h$ \cite{bipm}.  This one relates, through the relation $E=h\nu $ ($=\hbar \omega $), the energy $E$ of a photon to the frequency $\nu$  (or angular frequency $\omega =2\pi\nu$) of the corresponding electromagnetic wave. Fixing $h$ to a certain number of \,joule$\,\cdot\,$seconds, in agreement with its presently measured value  ($h \simeq 6.626\,... \times 10^{-34}$ J\,s\,, or $\hbar=h/2\pi \simeq  1.054\,...\times  10^{-34}$ J\,s) will allow us to define the unit of energy, the joule, as corresponding to a very large number of \,s$^{-1}$ ($ .948\,...\times 10^{34}$~s$^{-1}$).
The definition of the kilogram follows, such that 1 joule = 1~kg\,m$^2$\,s$^{-2}$ as usual, rendering obsolete the international prototype stored at BIPM. Of course the value of $h$ gets fixed in agreement with its present best determined value ($6.626\;070\;15\times10^{-34}~{\rm J\,s}$ \cite{codata}), so that the mass of the international prototype will still be practically equal to 1\,kg at the time of the change, within experimental uncertainties. The change  will not affect in any practical way the values of masses at the time it is performed, while allowing for  more precise measurements in the future.

\vspace{2.5mm}

We could choose to measure distances directly in seconds by fixing the speed of light at $c=1$, the natural choice in relativity; and energies and masses directly in s$^{-1}$ by fixing the value of  $\hbar = h/2\pi$  also at 1, the natural choice in quantum mechanics. This would be  conceptually much simpler.
But considering the second as being also a unit of distance, close to $3 \times 10^5$ km, and the second$^{-1}$ as a common unit  for energy and  mass (with $c=1$), would not be very practical. Indeed the energy unit  of 1~s$^{-1}$ would be very small, $\,1.054 \,...\times  10^{-34}$ joule. Furthermore, the kilogram being itself associated with a very large  energy $E=mc^2=  .898 \,...\times 10^{17}$~J, the s$^{-1}$ as a unit of mass would be comparatively even smaller, 
$\, (1.054 \,...\times  10^{-34}) /(.898 \,...\times 10^{17})$~kg \,i.e. $1.173 \,...\times 10^{-51}$ kg. This is also understood from $\hbar/c^2\simeq 1.173 \,...\times 10^{-51}$ kg~ which fixes the mass unit of 1 s$^{-1}$   at  $\,1.173 \,...\times 10^{-51}$ kg.

\vspace{2.5mm}

While such units are often very convenient in relativistic and quantum physics,
it would seem unpractical to replace square metres by $\,\approx 10^{-17}$ s$^{2}$ when measuring surfaces, or to ask for $10^{\hspace{.3mm}51}$\,s$^{-1}$ of potatoes on the market.
For historical and practical reasons we remain attached to measuring distances in metres rather than in seconds, and masses and energies in  kilograms  and joules rather than in s$^{-1}$. Fortunately we still have the freedom to define the metre, kilogram, and joule as  corresponding to fixed submultiples or multiples of the second, or second$^{-1}.$
This is how the units of length, and very soon energy and mass, are or will be derived from the unit of time by fixing $c$ and $\hbar$ to a certain number of m/s and J\,s according to $\,c=c_\circ$\,m/s, $\,\hbar=\hbar_\circ$\,J\,s, allowing to define the metre and the joule, 
and subsequently the kilogram, the newton, ...\,, from the second.
This is what the present redefinition of the SI, to become effective from 20 May 2019,  is going to achieve \cite{bipm}.

\vspace{2.5mm}
But we can still go further. Within relativity space and time, related by the Lorentz symmetry group, become quantities of similar nature, which may be measured with a common unit. It is the same for mass and energy.  Within quantum mechanics energies and angular frequencies are directly related, allowing us to measure energies in s$^{-1}$, as for 
angular frequencies.
In a conceptually ideal system we ought to choose $c=1$ as suggested by relativity, and $\hbar=1$ as suggested by quantum mechanics so that the intrinsic angular momentum, or spin, of the electron is simply 1/2. This is in fact implicitly supposed when we say that the electron is a spin-1/2 particle. Such a system, however, is usually viewed as an idealized one for theoreticians, requiring to move back to an ordinary system in which time, space, energies and masses are all measured  in terms of their own units, in seconds, metres, joules and kilograms, as in the traditional or newly revised SI.

\vspace{2.5mm}
We shall show here that the two points of view, although apparently antagonist, may be reconciled within {\em a single unified system including the new SI}, in which {\em both $c$ and $\hbar$ are also equal to 1}. \linebreak
Then {\em the metre, joule and kilogram get identified as specific submul\-tiples or multiples of the second or second\,$^{-1}$}, the natural units of length, energy, and mass in an ideal system with $c=\hbar=1$, once the second is chosen as the unit of time. We can thus have, simultaneously, 
\be
\label{hc}
\framebox [9.8cm]{\rule[-.6cm]{0cm}{1.5cm} $ \dis
\left\{\
\ba{cclcc}
c\!&=&\ \ 299\;792\;458 \ \,\hbox{m/s} \ \ &\hbox{\ \it \u{and\phantom p$\!\!\!$}}&\  \ \ c=1\,, 
\vspace{2mm}\\
h\!&=&\!6.626\;070\;15 \times 10^{-34}\  \hbox{J\,s} \ \ &\hbox{ \it \u{and\phantom p$\!\!\!$}}&\  \ \ \hbar=\dis {h}/{2\pi}=1 \,.
\ea
\right.
$}
\ee
These equalities will provide as desired the appropriate new SI definitions for the metre, the joule, and the kilogram, also made compatible with the advantages of working in a system with $\,c=\hbar=1$, in which
\linebreak  {\em the fundamental laws of physics do not even depend on these parameters}.  \,These were previously referred to as ``fundamental constants of nature''. Once identified with unity, they get downgraded to simply providing numerical  conversion factors between related units.

\vspace{2mm}

\section{The new system for the electrical units}
\label{sec:elec}

\vspace{1mm}

This new picture can be extended to the electrical units, up to now defined from the mechanical ones by fixing the value of the parameter $\mu_\circ$ defining the magnetic permeability of vacuum and entering the traditional definition of the ampere \cite{amp,amp2,SI}. The new approach we propose here remains applicable, as we shall see, in the context of the new SI  \cite{bipm}, in which the rigid connection between electrical and mechanical units gets somewhat loosened. This  is  a consequence of the recent decision taken at the last CGPM to define the coulomb (and thus the ampere and all electrical units) so that the numerical value of the elementary charge is exactly fixed, at $ \,e= 1.602\;176\;634 \times 10^{-19}$ C, requiring to turn $\mu_\circ$ into an adjustable parameter \cite{mise}. It is indeed the price to pay  for having decided to fix the numerical value of $e$, rather than measuring it as it has been done up to now.

\vspace{2mm}
Let us return to our new proposed approach. The present conventional choice $\mu_\circ=4\pi\times 10^{-7}$ N/A$^2$, or equivalently H/m, before the upcoming 2019 redefinition of electrical units,
may be reconciled with the ideal one $\u{\mu_\circ}=1$ 
by fixing the value of $\u{\mu_\circ}$, as done for $c$ and $\hbar$,
\vspace{-.5mm}
at a certain number of N/A$^2$. 
We shall thus write 
$\u{\mu_\circ}=\mu_\circ\,$\,N/A$^2$, 
where
$\mu_\circ$, now dimensionless, is  initially fixed to $4\pi\times 10^{-7}$. 
But, as from 2019 the coulomb will no longer be obtained from the traditional definition of the ampere \cite{amp,amp2,SI} but by fixing the value of the elementary charge \cite{bipm}, the value of  $\mu_\circ$ will still have be very slightly adjusted, from $4\pi\times 10^{-7}$ into  $\,\mu_\circ=4\pi\times 10^{-7}\,\eta^2$.

\vspace{2mm}

{\em In both cases\,}, and thus independently of the on-going 2018-2019 change for the definition of the electrical units,  we can ask for 
$\u{\mu_\circ}=1$, which provides an ideal system with $\hbar=c=\u{\mu_\circ}=\u{\epsilon_\circ}=1$.
\vspace{-1.1mm}
Fixing $\,\u{\mu_\circ} = \mu_\circ \, {\rm N/A}^2 = 1$  determines the ampere, which must verify
\be
\label{mua}
1\,{\rm A}^2 =\,\mu_\circ\, {\rm N}\,.
\ee
The ampere appears as {\em proportional to a square root of the newton}, which is conveniently expressed as  1\,A = $\sqrt{\mu_\circ\, \hbox{N}}$\,.

\vspace{1.5mm}

Altogether the set of three equations
\be
\label{hcmu9}
\framebox [8.5cm]{\rule[-.9cm]{0cm}{2cm} $ \dis
\left\{\ \ba{ccccccc}
c\!&=&\! c_\circ \ {\rm m/s}\!&=&\! 1\,,\!&&\!  \hbox{which fixes the metre,}
\vspace{2mm}\\
\hbar\!&=&\! \hbar_\circ \ {\rm J\,s}\!&=&\! 1\,,\!&&\!  \hbox{which fixes the joule,}
\vspace{1.5mm}\\
\u{\mu_\circ}\!&=&\! \mu_\circ \ {\rm N/A}^2\!&=&\! 1\,,\!&&\!  \hbox{which fixes the ampere}
\ea\right.
$}
\ee

\noindent
leads to a system that includes and goes beyond the present redefinition of the SI. It
allows us to reconcile, within a single system of  units, 

\vspace{1.5mm}
{\bf 1)} \  the advantages of a conceptually desirable system in which 
$\hbar=c=\u{\mu_\circ}=\u{\epsilon_\circ}=1$, 
the vacuum magnetic permeability, electric permittivity, and impedance being all equal to 1, with the elementary charge
$\,e = \sqrt{4\pi\alpha} = .3028\,...\,,
\,K_J=2e/h=e/\pi$ and $\,R_K=h/e^2= (376.730\,...\ \Omega)/2\alpha = 1/2\alpha $;
\vspace{1mm}

{\bf 2)} \
the convenience of our familiar SI units, second, metre, kilogram, joule, newton, ampere, coulomb, volt, etc. appropriately defined or redefined very much as usual;

\vspace{1mm}

{\bf 3)} 
\,with all these units also expressed in terms of the second, s$^{-1}$, s$^{-2}, \ ...\,$, \,or even as pure numbers as for the ohm, the coulomb, and the weber;

\vspace{1mm}

{\bf 4)} \
furthermore, as the numerical value of $e$ will also get fixed, the coulomb, the ampere and all electrical units get redefined accordingly, using an adapted  value of $\mu_\circ$ equal to $4\pi\times 10^{-7}\,\eta^2$ where $\eta$ is very close to 1; this adjustment of electrical units is automatically and explicitly taken care of \,by expressing the ampere and coulomb proportionally to $\sqrt{\mu_\circ}$, as indicated above.

\vspace{3mm}

With the values of $\hbar$ and $c$ getting fixed to $\,\hbar_\circ$\,J\,s and $\,c_\circ$\,m/s, respectively, we get the expression of the
\vspace{-3mm}
newton
\be
{\rm  1\,N\,=\,\frac{1\, J\,s}{1\,m/s}\ \,s^{-2}}\,=\,\frac{c_\circ}{\hbar_\circ}\ \,\rm s^{-2}\,.
\ee
This leads, most notably, to the following dual expressions for the ampere A and the coulomb C (at the moment still evaluated with $\mu_\circ=4\pi\times 10^{-7}$), the elementary charge $e$ and the impedance of the vacuum $Z_\circ$:

\vspace{-3mm}

\bea
\label{aez}
\left\{\ \ba{ccccccc}
1 \, {\rm A}&=&\sqrt{\mu_\circ \phantom{\hbox{\large I}} \!\!\rm N}&=&\dis \sqrt{\mu_\circ c_\circ/{\hbar_\circ }}\  
\ {\rm s}^{-1}&=&1.890\;067\;014\;853\,...\times 10^{18}\ \,{\rm s}^{-1} \,,
\\ [2.5mm]
1 \, {\rm C}&=&\sqrt{\mu_\circ \phantom{\hbox{\large I}} \!\rm kg\,m}&=&\dis 1/\sqrt{{\epsilon_\circ c_\circ \hbar_\circ}} \ \ \ \ \ &=&1.890\;067\;014\;853\,...\times 10^{18} \ ,\ \ \ \ \ \,
\\ [2.5mm]
e&=&1.602\;176\;634 \times 10^{-19} \ {\rm C} &=&\sqrt{4\pi\alpha}\ \ \ &\simeq&\,.302\;822\;1208\ ,\ \ \ \ \ 
\vspace{2.5mm}\\
Z_\circ&=&\mu_\circ c_\circ \ \Omega\,= \,376.\,730\;313\;461\,...\ \Omega\  &&  &=& 1\ . \hspace{28mm}
\ea\right.
\nonumber
\\ [-12mm]
 \\ [4mm] \nonumber
\eea

\vspace{-1mm}
\noindent
The elementary charge $e$ remains as {\em an independent dimensionless 
\vspace{.2mm}
free parameter equal to $\sqrt{4\pi\alpha}\,$, to be determined experimentally}.  However, it will get numerically fixed at exactly $ 1.602\;176\;634 \times 10^{-19}$ C \,in the new SI, at the price of suitably adjusting the coulomb and thus the value of $\mu_\circ$, now to be taken as $4\pi\times 10^{-7}\,\eta^2$.

\vspace{2mm}

The electron-volt, 1\,eV = 1.602\,...$\times \,10^{-19}$ J\,, gets exactly fixed at $(e_\circ/\hbar_\circ) \,  \rm s^{-1}\!=1.519\,... \times \,10^{15}\ \rm s^{-1}$. The volt, equal to $1/\sqrt{\mu_\circ c_\circ \hbar_\circ}\ \,\rm s^{-1}$,  is also recovered as 1\,eV/$e = 1.519\,... \times \,10^{15}\  \rm s^{-1}/.3028\,... = 5.017\,...\,\times \,10^{15}\  \rm s^{-1}$.
\,The Josephson and von Klitzing constants are related, respectively, to the size of the electron-volt and to the fine structure constant $\alpha$.  
They get fixed numerically in SI units (GHz/V and ohms) \cite{mise}
 thanks to the choice of $\,h=h_\circ$\,J\,s, $e=e_\circ$\,C. 
 At the same time they also become pure numbers equal to 
$e/\pi$ and $1/2\alpha$, respectively, 
\vspace{-.3mm}
thanks to our choice of $\,\hbar=1$ with $c=\u{\mu_\circ}=1$ so that 
$e=\sqrt {4\pi\alpha}$:
\be
\label{jk}
\hspace{0.5mm}\left\{\, 
\ba{ccccccccccc}
K_J\!\!&=&\!\dis {2e}/{h} &=&\dis {2e_\circ}/{h_\circ}\  \, \rm Wb^{-1}&= &483\;597.\,848\,416\,...  \  \rm GHz/V\!&=&\! {e}/{\pi} \!& \simeq & .0963\;912\;748\ ,
\vspace{2mm}\\
R_K \!\!&=&\! {h}/{e^2}&=&\dis \frac{h_\circ}{e_\circ^2}\ \Omega \,=\,\dis  \frac{\mu_\circ c_\circ\,\Omega}{2\alpha}\!&=  & \ 25\;812.\,807\,459\,... \  \Omega\! &=&{1}/{2\alpha }
\!&\simeq & 68.\,517\,999\,58\ .
\ea\right.
\ee
$K_J=e/\pi$ is equivalent to saying that 1\,eV\,= $\pi\, (2e_\circ/h_\circ) \, \rm s^{-1}=\pi \times .483\;597\,...\times 10^{15}
\ \rm s^{-1}=
1.519\,... \times \,10^{15}\  \rm s^{-1}$. 
The SI value in ohms of $R_K$ is the vacuum impedance, $Z_\circ=\mu_\circ c_\circ \,\Omega = 376.730\,313\,...\, \Omega$ (now also equal to 1), multiplied by $137.\,035\,999\,...\, /2\,$, providing the well-known 
$\,25\,812.\,807\,...\ \Omega$\,. It is now, at the same time, also dimensionless and equal to 137.\,035\,999\,...\,/2.

\begin{table}
\caption{\ 
The new dual system\hspace{-.5mm}:
\vspace{-.5mm}
$\,c=c_\circ\,{\rm m/s}=1$ and $\hbar=\hbar_\circ\,{\rm J\,s}=1$ 
fix the metre and the joule. \,$\u{\mu_\circ}= \mu_\circ\, {\rm N/A}^2=1$ allows us to express 
\vspace{-.3mm}
the ampere and coulomb  as  \,1\,A = $\sqrt{\mu_\circ\,{\rm N}}$ and 1\,C =  $\sqrt{\mu_\circ\, {\rm kg\,m}\,}$.
\vspace{-.1mm}
\,With $\hbar/c=(\hbar_\circ / c_\circ) $\,kg\,m = 1, \,the coulomb, dimensionless,  is 
\vspace{-.4mm}
equal to
$\sqrt{{\mu_\circ c_\circ}/{\hbar_\circ }}= $ $1.890\,... \times 10^{18}$.
\vspace{.1mm} 
The elementary charge is $e=1.602\,... \times 10^{-19}$\,C $ = \sqrt{4\pi\alpha}= .3028\,...\,$, and 
\vspace{.1mm}
the impedance of vacuum $\,Z_\circ=\mu_\circ c_\circ \,\Omega=$  $376.730\,... \ \Omega = 1$, with $K_J\!=2e/h=e/\pi$ and $\,R_K\!=h/e^2=1/2\alpha\,$.  The dimensionless coulomb, ohm and weber are related by 1\,C\,$\times \,1\,\Omega$ = 1\,Wb\,.
\vspace{1mm}\\
{\hspace*{2mm}}
The left column refers to the units and remarkable quantities expressed in a conventional way in
terms of the s, m and kg, 
with the electrical units derived from the ampere expressed as  $\sqrt{\mu_\circ\rm N}$. The two other columns give their expressions in s, s$^{-1}$, s$^{-2}$ or as dimensionless constants. This is the case for the coulomb, ohm, and weber, and for $e=\sqrt{4\pi\alpha}$, the impedance of vacuum $Z_\circ =1$, and flux quantum $\Phi_\circ =\pi/e$. 
The numerical values are evaluated here for $\mu_\circ= 4\pi\times 10^{-7}$, 
\vspace{-.5mm}
and the corresponding benchmark value $\,\alpha=\alpha_\circ= 1/137\,035\,999\,158\,713\,...$ as in (\ref{alphac}).  
They should be rescaled using the  parameter  $\eta=\!\sqrt{\alpha/\alpha_\circ}\,$ according to   (\ref{varelec}), and may change by a few $10^{-10}$, depending on $\alpha$.
\vspace{2mm}
\label{tab:unit}
}
\begin{tabular}{c}
$ 
\ba{|ccccc|}
\hline
&&&& \\ [-3mm]
 &&&&  \hbox{\hspace{-30mm}Expressions in terms of s, s$^{-1}$, s$^{-2}$ or as constants\ \ \ }
 \\ [-2mm]
 \hbox{\ \ ``Conventional expressions''}
 &&&& \\ [-2mm]
&&&& \hspace{-30mm}(\,\hbar=c=\u{\mu_\circ}=\u{\epsilon_\circ}=1)
\\ [2.5mm]
\hline
&&&& \\ [-2mm]
\ \ 1 \ \hbox{m}\!\!&=&\ \dis\frac{1}{c_\circ}\ \ {\rm s}\ \ \ &=& \dis\frac{1}{299\;792\;458}\ \ \hbox{s}\ \ 
\\ [3.5mm]
\ \ 1\ \hbox{J}\!\!&=& \,  \dis \frac{1}{\hbar_\circ} \ \, \hbox{s}^{-1}
&=&\dis
\ \, .948\;252\;156\;246\,... \times 10^{34}\ \ \hbox{s}^{-1} \ \  
 \\ [3.5mm]
 \ \  1 \ \hbox{kg} \!&=& \dis \frac{c_\circ^2}{\hbar_\circ} \ \,\hbox{s}^{-1} &=&\dis
\ \, .852\;246\;536\;175\,... \times 10^{51}\ \ \hbox{s}^{-1} \ \ 
  \\ [3.5mm]
\ \ \ 1\,{\rm N}
&=&
\dis \frac{c_\circ}{\hbar_\circ}  \ \,\hbox{s}^{-2} &=&\dis
  2.842\;788\;447\;250\,... \times 10^{42}\ \ \hbox{s}^{-2} \ \ 
 \\ [4mm]  \hline
&&&&  \\ [-2mm]
\ \     1 \ {\rm A} = \sqrt{\,\mu_\circ\,{\rm N}}&=& \dis  \,\sqrt{\,\frac{\mu_\circ c_\circ}{\hbar_\circ }}\  \ {\rm s}^{-1}
&=&  1.890\;067\;014\;853\,...\times 10^{18}\ {\rm s}^{-1}  \ \ 
\\ [3.5mm]
\ \ \ \ \ \ 1 \ {\rm C}=\sqrt{\,\mu_\circ\ \rm kg\,m} &=& \dis  \,\sqrt{\,\frac{\mu_\circ c_\circ}{\hbar_\circ }}\  \ \ \ \ \  &=&   1.890\;067\;014\;853\,...\times 10^{18}\ \ \ \ \ \ \ \,
\\ [3.5mm]
\ \  \ \ 1 \ {\rm V} = 1 \ {\rm J/C}\ \ \ \ &=&   \dis
\frac{1}{\sqrt{\mu_\circ c_\circ \hbar_\circ}}\  \ {\rm s}^{-1}&=&  5.017\;029\;284\;119\,...  \times 10^{15}\ {\rm s}^{-1}\ \ 
\\ [4mm]
\ \ \ \ \ 1\,\hbox{V/m}\,=
\, \ \ 1\,{\rm N/C} \  \ \
=\dis \frac{1}{\sqrt {\mu_\circ}} \,\sqrt{\hbox{J/m}^3}\ \rm m/s \!&=&\! \dis \sqrt{\,\frac{c_\circ}{\mu_\circ\hbar_\circ}}\ \ {\rm s}^{-2}
\!&=&\dis  1.504\;067\;540\;944\,... \times  10^{24}\ {\rm s}^{-2}\ \ 
\\ [3.5mm]
\ \ \ 1\,\hbox{T}\ =
1\,{\rm N/(A\,m)}
=\dis \frac{1}{\sqrt {\mu_\circ}} \,\sqrt{\hbox{J/m}^3}\!&=&\! \dis \sqrt{\,\frac{c_\circ^3}{\mu_\circ\hbar_\circ}}\ \ {\rm s}^{-2}
\!&=&\dis  4.509\;081\;050\;976\,... \times  10^{32}\ {\rm s}^{-2}\ \ 
\\ [4mm]
\ \ 1\ {\rm Wb \,=\, 1 \ V\,s }\ \ \ \ &=&\!\dis
\frac{1}{\sqrt{\mu_\circ c_\circ \hbar_\circ}}\!&=& 5.017\;029\;284\;119\,...  \times 10^{15} \ \ \ \ \ \ \ 
\\ [-.5mm]
..............................................................................
&
\!\!\!\!\!.........
&
\!\!\!\!..........................
&
\!\!\!\!........
&
\!\!\!\!......................................................
\\ [1mm]
\ \ \ \,1 \ \hbox{F}\ =\ \,\dis 1\,{\rm C/V\ \,=\ 1\,s/\Omega\, =\ \mu_\circ\, s^2/m }
&=&\ \ \mu_\circ c_\circ\ {\rm s}\ \ 
\!&=&\! \ \ \ 376.730\;313\;461\,...\ {\rm s}
\\ [2mm]
\,1\  \hbox{H}\ =\
1\ \hbox{J/A}^2\ =\ 1\ \Omega\,{\rm s}\ =\  \dis \frac{1}{\mu_\circ }\ {\rm m} &=&\dis\  \frac{1}{\mu_\circ c_\circ}\ {\rm s} \ \ 
\!&=&\!\dis 1/376.730\;313\;461\,...\ {\rm s}
\\ [3mm]
\ \ \ \  1 \ \Omega \ =\, 1 \ {\rm V/A}= 1 \ {\rm W/A}^2=\, \dis \frac{1}{\mu_\circ }\ {\rm m/s}
\ &=& \dis \frac{1}{\mu_\circ c_\circ }\ &=&  1/376.\,730\;313\;461\,...\, \ \  \
\\ [2.5mm]
\ \ \ \ \dis Z_\circ\,=\,\mu_\circ c_\circ \ \Omega\,= \,376.730\;313\;461\,...\ \Omega\  &=&  \u{\mu_\circ} \,c \
  &=&   1
\\ [3mm]
 \hline
&&&&  \\ [-2mm]
\ \ \ \ \ \  \ \   e\ \,=\, 1.602\;176\;634 \times 10^{-19} \ {\rm C} &=&\sqrt{\,4\pi\alpha}\ \ \  &=&.302\;822\;120\;789\,...\ \ 
 \\ [3mm]
\ \  \ \  1\ {\rm eV}\,=\,  1.602\;176\;634 \times 10^{-19} \ {\rm J}&=& \dis
\ \ \ \ \  \frac{e_\circ}{\,\hbar_\circ}\ \ \ {\rm s}^{-1}  &= & \!\! 1.519\;267\;447\,878\,... \times 10^{15}\ {\rm s}^{-1}\ \
\\ [2.5mm]
\ \ \Phi_\circ =\,h/2e \,= \,2.067\;833\;848\;461\,... \times 10^{-15}\  {\rm Wb}\!\!\!    &=& 
\dis  \sqrt{\,\frac{\pi}{4\alpha}}\,=\,\frac{\pi}{e}\  \ &= &\ \ \ 10.374\,382\,972\,...\ \ \ \ \ \ \ \ \ \ \ \ \ \ 
\\ [3mm]
\ \ \ \ K_J =\,2e/h\, = \,483\;597.\,848\;416\;983\,... \  {\rm GHz/V}    &=& 
\dis  \sqrt{\,\frac{4\alpha}{\pi}}\,=\,\frac{e}{\pi}\  \ &=& .0963\;912\;748\;023\,...\, 
\\ [3mm]
\dis R_K \,=\, \frac{h}{e^2}\,=
\, \frac{\mu_\circ c_\circ\, \Omega }{2\alpha}\,= \,25\;812.\,807\;459\;304\,...\  \Omega \!\!\! \! \!\!&=&  
\dis \frac{\u{\mu_\circ}c}{2\alpha} \,=\, \frac{1}{2\alpha}&= & 68.517\,999\,579\,...\ \ \ \ \ \ \ \ \ \ 
\\ [5mm]
\hline
\ea 
${}
\end{tabular}
\end{table}

\vspace{2mm}
The resulting expressions of the various mechanical and electrical units are given in  Table \ref{tab:unit}.
We leave aside for the time being the question of a more fundamental definition for the unit of time, that one might like to obtain in the future from  the electron mass $m_e$ using $\,\hbar/m_e c^2 \simeq  1.288\;088\;667 \times 10^{-21}$\,s.

\section{The implications of fixing the value of the elementary charge}
\label{sec:should}

Let us discuss further the consequences  of fixing exactly, in the planned revision  of the SI to become effective from 20 May 2019 \cite{bipm}, the numerical value of the elementary charge $e$ when expressed in coulombs,
according to the additional 
requirement 
\be
\label{ee0}
e\,=\,e_\circ\, {\rm C} \,=\, 1.602\;176\;634 \times 10^{-19}   \     {\rm C}\,.
\ee
Although this may be conceptually questionable, this is intended to provide a very precise way to fix the coulomb, and the other electrical units, allowing for more precise measurements. 

\vspace{2mm}
However, adopting the new expression (\ref{ee0}) of the elementary charge $e$ requires adjusting suitably the size of the coulomb, at present defined as 1\,A\,s, and thus the size of the ampere.
Writing the two expressions of $e$ according to the old and new definitions of the coulomb,
$
\,e=e_{\rm old}\, {\rm C}_{\rm old}=e_\circ \, \eta \ {\rm4\pi\times 10^{-7} C}_{\rm old}=e_\circ\, {\rm C}
$,
we see that the redefined coulomb C may be ``larger'' (or possibly smaller) than the earlier one by a factor
\be
\label{varc}
\eta\,=\,\frac{\rm C\ \,}{{\rm C}_{\rm old}}=\frac{e_{\rm old}}{e_\circ}\ ,
\ee
very close to 1\,.
The coulomb, and thus the ampere, equal to $\sqrt {\mu_\circ\, \rm N}$, get multiplied by $\eta$, with
\be
1\,{\rm A}  = \sqrt{\mu_\circ\, \rm N}\, =\, \sqrt{\,4\pi\times 10^{-7}\ {\rm N}}
\, \times \,\eta\ ,
\ee
so that $\mu_\circ$ should be multiplied by $\eta^2$, becoming $\mu_\circ=4\pi\times 10^{-7}\,\eta^2$.

\vspace{2mm}

We must ensure that the two concurrent definitions of the coulomb, and the ampere, are compatible.
The ampere is expressed  as 1\,A = $\sqrt{\mu_\circ \,\rm N}$  as in (\ref{aez}).  
\vspace{-.5mm}
The compatibility of the set of 4 equations in (\ref{hcmu9},\ref{ee0}) requires that the third equation in (\ref{hcmu9}),
$\,\u{\mu_\circ}=\mu_\circ \, {\rm N/A}^2 = 1$,
\,be used, 
\vspace{-.5mm}
no longer to fix the exact size of the ampere from a fixed 
$\mu_\circ = 4\pi\times 10^{-7}$, but instead to determine a new ``floating'' value for $\mu_\circ$, from now on equal to $4\pi\times 10^{-7}\,\eta^2$, from an ampere already defined as 1\,C/s.
The two definitions for the coulomb (and ampere) from (\ref{aez}) and (\ref{ee0}) should be equalized, as follows
\be
\label{equiv}
\framebox [14.2cm]{\rule[-.35cm]{0cm}{.9cm} $ \dis
\ 1\,{\rm A} = \sqrt{\mu_\circ \phantom{\hbox{ I}} \!\!\!\rm N}  
\,=\,\sqrt{\mu_\circ c_\circ /{\hbar_\circ }}\   {\rm s}^{-1}\ \ \ \Longrightarrow\ \ \ \ 1 \ {\rm C}\,=\,1/\sqrt{{\epsilon_\circ c_\circ \hbar_\circ }}\, =\,{e}/{ e_\circ}\ \ \ \ \Longleftarrow \ \ \ e=e_\circ\, \rm C\ \ . \, 
$}
\ee
This provides two equivalent expressions for the fine structure constant,
\be
\label{2alpha}
\alpha\,=\,\frac{e_\circ^2}{4\pi\epsilon_\circ\hbar_\circ c_\circ}\ = \ \frac{\mu_\circ c_\circ\,e_\circ^2}{4\pi\hbar_\circ }\,=\,\frac{e^2}{4\pi} \ .
\ee
$\alpha=e^2/4\pi$ corresponds  to the choice $\,\hbar=c=\u{\mu_\circ}=\u{\epsilon_\circ}=1$ also made here, leading to the dual expression of the elementary charge
\be
\label{ez}
e\,=\,e_\circ\, {\rm C} \,=\, 1.602\;176\;634 \times 10^{-19}   \     {\rm C}\,=\,\sqrt{4\pi\alpha}\,\simeq\,.302\;822\;1208\,.
\,\ee

\vspace{2mm}

The price to pay for fixing $e_\circ$ as above
 is that  $\mu_\circ$ should now be multiplied by $\eta^2$, getting adjusted  to 
\be
\label{mualpha}
\mu_\circ=\,\frac{4\pi \hbar_\circ}{c_\circ e_\circ^2}\ \alpha 
\,=\,4\pi\times 10^{-7}\ \eta^2\ ,
\ee
so that $\alpha$ stays unchanged.
This leads to evaluate the benchmark value for $\alpha$ corresponding to keeping an unchanged  $\mu_\circ=4\pi\times 10^{-7}$ with the new fixed choice of $e_\circ$, 
\be
\label{alphac}
\alpha_\circ\,=\dis \,\frac{4\pi\times 10^{-7}\,c_\circ\,e_\circ^2}{4\pi\,\hbar_\circ}\,=\,7.297\;352\;565\;305\,...
\times 10^{-3}\,=\,1/137.\,035\;999\;158\;713\,...\,.
\ee
It is very close to the present best-determined value $\alpha=1/137.\,035\;999\;139\,  (31)$, with a relative standard uncertainty of $2.3 \times 10^{-10}$ \cite{alpha,codata}.
\,It corresponds to
\be
\label{alphazero}
\sqrt{4\pi\alpha_\circ}\,=\,.302\;822\;120\;789\;201\,...\ .
\ee

$\mu_\circ$ is no longer rigidly fixed, but proportional to $\alpha$.  
Fixing $e_\circ$ as in  (\ref{ee0},\ref{ez}) leads to a very small rescaling of all electrical units depending on 
the parameter 

\vspace{-3.5mm}
\be
\label{etaz}
\eta\,=\,\sqrt{{\alpha}/{\alpha_\circ}}\,\simeq \,1\ \ \ (\hbox{up to $\,\simle\,$ a few} \ 10^{-10}\,)\,.
\ee

\vspace{-3mm}

\noindent
Thus 

\vspace{-5mm}
\be
\hbox{fixing \ $e=e_\circ$\,C}\ \ \ \Longrightarrow\ \ \hbox{\em all electrical units become dependent on $\alpha$\,,}
 \ee
 \vspace{-5mm}

\noindent
with

\vspace{-5mm}

 \bea
 \label{varelec}
 \framebox [15.7cm]{\rule[-.3cm]{0cm}{.8cm} $ \dis
 \rm (A,C) \propto \eta\,,\ \,  (V,T,{\rm Wb}) \propto \eta^{-1},\ \, F\propto \eta^2,\ \, (H, \Omega) \propto \eta^{-2}\,;\ \ \  \mu_\circ\propto \eta^2,\ \, \epsilon_\circ \propto \eta^{-2}\,; \ \ \hbox{with} \ \ \eta=\sqrt{\alpha/\alpha_\circ}\ .
 $}
 \nonumber\\ [0mm] 
 \eea
 
 \vspace{-2mm}
 
 \noindent
 The sizes of the electrical units themselves become dependent on future experimental measurements of the fine structure constant $\alpha$.
 \vspace{2mm}
 
In particular the value in ohms of the ``impedance of the vacuum'', $\mu_\circ c_\circ\propto \alpha$, now depends on $\alpha$ in the new SI,
i.e. on what was previously the value of the elementary charge $e$ ! 
Of course  $Z_\circ$ itself, equal to $\mu_\circ c_\circ\,\Omega$, does not, as one can verify  from (\ref{varelec}). Still this artificially introduced dependence of the measure of the impedance of vacuum
may look strange, especially for a quantity that is basically 1, with an appropriate choice of fundamental units (see below, subsection \ref{sub:imp}).
Its measure {$\mu_\circ c_\circ$ should actually be viewed as  a parameter characterizing the size of the ohm (unfortunately dependent on the experimentally measured value of $\alpha$ in the new SI), and not as a measure of a physical property of the vacuum. 

\vspace{2mm}
 
This  conceptual inconvenience of the new SI associated with the floating character of the electrical units gets 
alleviated within the unified framework proposed here, in which the impedance of  vacuum, $Z_\circ=\mu_\circ c_\circ\,\Omega$, while still equal to 376.730\,...\ $\Omega$, is also identical to 1. The magnetic permeability of vacuum, now expressed as $\u{\mu_\circ}\,=\mu_\circ \rm \ N/A^2=1$, remains also equal to 1 with 1\,A = $\sqrt{\mu_\circ \rm N}$, 
\vspace{-.7mm}
even 
if the ampere A, as well as the numerical value  $\mu_\circ$ of the vacuum permeability (expressed in N/A$^2$ or H/m), are both dependent on $\alpha$, as seen in (\ref{varelec}).

\vspace{2mm}

In case  $\alpha$ were to vary with time as the possibility is occasionally  considered, {\it all electrical units, including the dimensionless ohm, coulomb, and weber,  would also vary with time.} This may be expressed through the set of equations
\be
\frac{\dot{\rm A}}{\rm A}\ = \  \frac{\dot{\rm C}}{\rm C}\ = \,-\,\frac{\dot{\rm V}}{\rm V}\ =\, -\,\frac{\dot{\rm T}}{\rm T} \ = 
\, -\,\frac{\dot{\rm Wb}}{\rm Wb} \ = \ \frac{1}{2}\ \frac{\dot\alpha}{\alpha}\ ,\ \ \ \ \ \ \ 
\frac{\dot{\rm F}}{\rm F}\ = \,-\,\frac{\dot{\rm H}}{\rm H}\ =\, -\,\frac{\dot{ \Omega}}{\Omega}\ 
= \ \frac{\dot\alpha}{\alpha}\ ,
\ee
as seen from eqs.\,(\ref{varelec}) and Table \ref{tab:unit}.
 This is  a not-so-attractive consequence of the new SI choice of fixing $e_\circ$ as in (\ref{ee0}), in principle not compatible with the requirement that units should not change with time.
 Fortunately there are very strong limits on a possible time variation of $\alpha$ (with $|\dot\alpha/\alpha |$ constrained to be at present $ < 10^{-16}/ \rm y)$, so that in practice this does not appear as a limitation.

\vspace{2mm}
Let us now return for some time to the mechanical units, defining the metre, the joule, and the kilogram by fixing the numerical values of $c$ and $\hbar$.

\ghost{
\vspace{4mm}
\bc
*\hspace{8mm}*
\vspace{4mm}
*
\ec
}

\section{\boldmath Defining the metre from the second, by fixing $\,c$}

The unit of time, the second, has  long been defined as the fraction 1/86\,400 of the ``mean solar day''. It was then redefined from a physical phenomenon involving the period of a specific atomic transition of the caesium-133 atom. More precisely,  since 1967 \cite{sec},
\be 
\ba{c}
\hbox{\it ``\,The second is the duration of \,9\,192\,631\,770 periods of the radiation corresponding to the transition}
\vspace{.1mm}\\
\hbox{\it between the two hyperfine levels of the ground state of the caesium-133 atom.''}
\vspace{-4mm}\\
\ea
\ee

\vspace{3mm}

The unit of length, the metre, first introduced as  ``la dix-millioni\`eme partie du quart du m\' eridien terrestre'' \cite{metre}, was  realized as the ``M\`etre des Archives'' in 1799, then defined from 1889 to 1960 as the length of a specific object, the international prototype of platinum-iridium stored in the ``Pavillon de Breteuil'' at S\`evres.
It was replaced in 1960 by a definition involving the wavelength of a krypton-86 radiation \cite{kryp}.
The theory of relativity, based on the invariance of the speed of light in vacuum, $c$,  allowed  for a new definition of the metre by relating it to the unit of time. This was done in 1983 by fixing the value of $c$ at, {exactly},
$ 
c= 299\;792\;458$ m/s \cite{c}.
Since this 
\vspace{-1mm}
date
\be 
\label{m}
\ba{c}
\hbox{\it ``\,The metre is the length of the path travelled by light in vacuum}
\vspace{.1mm}\\
\hbox{\it during a time interval of 1/299\;792\;458 of a second.''}
\ea
\ee
This is now reformulated, in an equivalent way, by stating that \cite{bipm}
\be
\label{c2}
\hbox{\em 
``\,the speed of light in vacuum $c$ is \,299\;792\;458 {\rm m/s}''}.
\ee

\vspace{1.5mm}

Indeed, relativity relates intimately  the concepts of time and space,
 describing events in a 4-dimensional spacetime with coordinates 
 $ x^\mu=\,(ct,\,\vec x)$.
It also relates the energy $E$ and momentum $\,\vec p\,$ of a particle of mass $m$ into the components of a 4-vector
$p^\mu\!= (E/c,\,\vec p\,)$,
with $\,p_\mu p^\mu \!=E^2/c^2-\vec p\,^2\!=m^2 c^2$. This reduces to the famous $E=mc^2$ for a massive particle at rest, and $E=pc$ for a massless photon travelling at the speed of light.
The unit of length may thus be derived from the unit of time by fixing the value of $c$.

\section{\boldmath  Fixing  $\,c=$ 299\;792\;458 \lowercase {m}/\lowercase{s}\,, \,or $\,c=1\,$, \,or both at the same time}

Choosing $c=1$ would be  the most natural choice, leading to measure space distances directly in seconds,
and energies and momenta in units of mass. 
The unit of length $u_l$ would then be  the distance travelled by light during the unit of time $u_t$. 
Once this one is defined as the second, $u_l$ is the length travelled by light in vacuum during 1\,s (sometimes called a light-second), namely in SI units
\be
\label{ul0}
\hbox{\em unit of length}\ u_l\,=\,c \ {\rm s}\,= \, 299\;792\;458 \ \hbox{m}\,.
\ee
This  is almost the distance to the Moon.
Choosing $c=1$ could  lead to aban\-don the metre to replace it by a  unit of length almost 300 million times larger, which is usually rejected as unpractical.

\vspace{2mm}

In spite of that, we remain free to continue using the metre, now defined as a submultiple of the above natural unit of length according to
1\;m = 1/299\;792\;458\ \,$u_l$\,.
This expresses that the speed of light  is chosen to be
$
c= 1 \,u_l/\,1\,u_t\,=\,299\;792\;458$ m/s.

\vspace{2mm}

Even better, we can do both things at the same time by requiring $\,c= 299\;792\;458 \ \hbox{m/s}\,$ and, ``en m\^eme temps'', 
$c=1\,$. We can  then complete the official formulation (\ref{c2})  as follows
\be
``\,\hbox{\em 
the speed of light in vacuum is \  c = 299\;792\;458\  {\rm m/s}\ \ $\equiv$\ 1\,"}.
\ee

\vspace{1mm}

\noindent
It expresses that the natural unit of length according to relativity, $u_l= 299\;792\;458$\;m, is not inde\-pendent from the unit of time, and may be identified with it.  We then
get 
\be
\label{mbis}
1 \ \hbox{m}\,=\,\frac{1}{299\;792\;458}\ \,\hbox{s}\,.\,
\ee
Expressing that $c=c_\circ$ m/s $\equiv$ 1 allows for the identification 1 m $=(1/c_\circ)$ s.

\vspace{2mm}

This unconventional formulation defines directly the metre as a submultiple of the second. It may be taken as an improved definition of the metre, 
reconciling the present one in (\ref{m},\ref{c2}) with the choice $c=1$ suggested by relativity. Conversely, the metre 
 also appears as a unit of time, as  is natural owing to the symmetry between space and time provided by relativity. We may then say that
\be 
\label{m2}
\ba{c}
\hbox{\em `` The time interval taken by light to travel in vacuum along a path of one metre long}
\vspace{.1mm}\\
\hbox{\it is \,1/299\;792\;458\,  of a second, or ...  one metre.''}
\ea
\ee

\noindent
We may even define a common subunit for distance and time appearing as a (metric) foot, equal to
1\,ns\, =\, .299\;792\;458 m $\,\simeq$ \,29.98\ cm,
again illustrating that the speed of light is $\,c=1$.

\vspace{2mm}

We still need a unit of mass. This one is defined, since the first ``Conf\'erence G\' en\' erale des Poids et Mesures''  in 1889,
 as the mass of the international prototype of the kilogram (IPK),  also known as ``le grand K'', {\boldmath $\cal K$}, conserved at the BIPM.  This was formulated in 1901 as \cite{kilo}:
\be
\!\! \hbox{\em ``The kilogram is the unit of mass\hspace{-.1mm}; it is equal to the mass of the international prototype of the kilogram.\hspace{-.3mm}''}
\ee
The unit of force is then  the newton, equal to \hbox{1~kg\,m\,s$^{-2}$}, and the unit of energy the joule, equal to 1~N\,m = 1~kg\,m$^2$\,s$^{-2}$. The unit of angular momentum, or action, is the kg\,m$^2$\,s$^{-1}$, or J\,s.
But the quantity of matter of the international prototype of the kilogram cannot remain exactly constant, and varies very slightly over the years.  In fact the differences in mass between the IPK and supposedly identical  copies average to a few tens of $\mu$g per century \cite{ipk}.
Could we avoid having to resort to such a physical object to define the unit of mass, and  relate it instead 
to a reproducible universal pheno\-menon, as for the second and the metre? 
This is where quantum physics comes in, as angular momenta and actions are quantized in units of $\hbar=h/2\pi$,
 whose value in SI units is $\,\simeq \, 1.054 \,...\times 10^{-34}$ J\,s\,.

\section{Quantum physics and the correspondence principle}

Quantum physics allows to  relate energies with frequencies, and momenta with wavelengths. It involves the Planck constant $h$, which relates, through
$
E=h \nu= \hbar \omega\,,
$
the energy $E$ carried by a photon with the frequency $\nu=\omega/2\pi$ of the corresponding electromagnetic wave, 
\vspace{-.3mm}
and similarly for the other particles. \linebreak A particle of momentum $\,\vec p\,$ is associated with a wave of wave vector $\vec k\,$
 given by
$
\vec p \,=\, \hbar \vec k\,,
$
and wavelength $\lambda$ given by the De Broglie relation
$
\,p = \hbar k= {h}/{\lambda}\,,
$
where $\hbar=h/2\pi$ is the reduced Planck constant.

\vspace{2mm}

Quantum mechanics thus relates, through the constants $h$ or $\hbar$, energies with inverses of times, and momenta with inverses of lengths,
$E \propto t^{-1},\ p \propto l^{-1}$.
This is an expression of the {\it correspondence principle}, 
which associates with each particle (including the photon as a quantum of light) of energy $E$ and momentum $\vec p\,$ 
\vspace{-2mm}
a wave
\be
\label{cor1}
\psi(\vec x, t)\,\propto\, e^{i\,(\vec k.\vec x-\omega t)} =\, e^{-\,i\,(Et-\vec p.\vec x)/\hbar}\,, \ \ \hbox{with}\ \ \ E=h\nu=\hbar \omega\,,\ \ \vec p\,=  \hbar \vec k \,  .
\ee
It may be formulated in a Lorentz-invariant way in terms of the spacetime coordinate $x^\mu=(ct,\vec x)$, as 
\be
\label{cor2}
\psi(x^\mu)\, \propto\,e^{-\,i\,k_\mu x^\mu} \!\!= \,e^{-\,i\,p_\mu x^\mu/\hbar}\,, \ \ \ \hbox{with}\ \ 
p^\mu\!= (E/c,\,\vec p\,)
=\hbar k^\mu\!=\hbar\  (\omega/c,\,\vec k\,)\,.
\ee
Thus a measurement of energy, or momentum, can be replaced by an associated measurement for the corresponding angular
frequency, or wave number, once the numerical value of $\hbar$ (in J\,s) is experimentally determined.  
For a free particle of mass $m$, $\,p_\mu x^\mu\!=mc^2\,\tau$ where $\tau$ is the proper time expe\-rienced by the particle, so that  $\psi(x^\mu)$  in (\ref{cor2}) 
\vspace{-.1mm}
may also be expressed as a function of the proper time, proportionally to $e^{-i\,mc^2\tau/\hbar}$.
\vspace{2mm}

Even better, a measurement of energy (or momentum) becomes identical to a measurement of angular frequency (expressed in s$^{-1}$) or wave number (expressed in m$^{-1}$), if the value of $\hbar$ gets fixed, in agreement with its present best experimental determination, within uncertainties. Units of energies  or momenta get then derived from the corresponding units of angular frequencies or wave numbers. The present definition of the kilogram, taken since 1889 as the mass of the international prototype stored at the BIPM, is no longer necessary, and may be abandoned in favor of a new definition based on quantum physics.

\vspace{1mm}

\section{Defining the \,joule\, and {\,kg\,} \vspace{1.5mm}from the inverse of the second, \hbox{ through quantum mechanics}}

\vspace{-1mm}

$\hbar$ is the fundamental quantum of action, or angular momentum,  in quantum mechanics. It has the dimensions of \,(energy $\times $ time), or
\,(momentum $\times$ space), or \,angular momentum ($\times$  angle, dimensionless). It is measured in J\,s, or equivalently kg\,m$^2$\,s$^{-1}$, as also seen from the expressions of the operators hamiltonian, momentum, and orbital angular momentum, 
\be
H=\,i\hbar\ \frac{\partial\ }{\partial t}\,,\ \ \ \vec P= -\,i\hbar\ \frac{\partial\ }{\partial \vec x}\,,\ \ L_z= -\,i\hbar\
\frac{\partial\ }{\partial \varphi}\ ,
\ee

\vspace{-1mm}
\noindent
in agreement with the correspondence principle. 
\vspace{-.4mm}
 In particular the plane waves 
$\psi\propto \, e^{i\,(\vec k.\vec x-\omega t)}$ in (\ref{cor1}) are eigenfunctions of $H$ and $\vec P$ with the eigenvalues $E=\hbar\omega$ and $\vec p=\hbar \vec k$\,. 
\vspace{.1mm}
They are also eigenfunctions of the operator $\,i\frac{\hbar}{c^2} \frac{d\ }{d\tau}$ where $\tau$ is the proper time experienced by the particle, with its mass $m$ as eigenvalue.

\vspace{2.5mm}

We make no difference between the so-called ``inertial''  and ``gravitational''  masses of a particle or macroscopic object, $m_i$ and $m_g$. They may be taken as identical, as investigated  long ago by E\" otv\" os and his collaborators. 
\cite{eot}. 
The resulting Equivalence Principle, formulated by Einstein, is at the basis of General Relativity, which provides the theory of gravitation, at the classical level. The {\em MICROSCOPE\,} experiment provides at present the most stringent test on the validity of this principle, allowing for the identification of inertial and gravitational masses at a level of precision of $\,2\times 10^{-14}$, \,for the pair of materials tested~\cite{touboul}.

\vspace{2.5mm}

The reduced Planck constant, $\hbar=h/2\pi$, is such that
the intrinsic angular momentum of the electron, or spin,  is $\hbar/2$. The electron is a spin-1/2 particle, 
meaning that its spin is  $S=$~1/2, when $\hbar$ is taken equal to 1\,.
Fixing $\hbar=1$ also defines the unit of energy from (or even identical to) the unit of angular frequency, namely the s$^{-1}$ (in principle $\times\  \hbar$, which is 1).
The expression of $\hbar$ (close to $1.054\;571\;818   \times 10^{-34} $ J\,s\linebreak if $\,h=h_\circ$  J\,s is fixed at 
6.626\;070\;15 $ \times 10^{-34}$ J\,s\,), implies that the natural unit of 
energy is
\be
\label{ue}
\hbox{\em unit of energy\,}\ u_e = \, \hbar \ \hbox{s}^{-1}=\,\hbar_\circ\ {\rm J}=\dis\, \frac{6.626\;070\;15 \times10^{-34}}{2\pi}\ \, \hbox{J}\
=\  1.054\;571\;817\;646 \,... \times \,10^{-34}  \ \,\hbox{J}\ \ 
\ee
(or kg\,m$^2$\,s$^{-2}$). Fixing $c=1$  also determines, as in (\ref{ul0}), the unit of length $\,u_l=c\;{\rm s} =c_\circ$\;m\, as the distance travelled by light in vacuum during the unit of time, here the second.
This ultimately allows for identifying the unit of length with the unit of time, and thus to measure distances directly in seconds.

\vspace{2.5mm}

The unit of mass is then the same as for energy (as $c=1$), namely $u_m=u_e= 1 $ s$^{-1}$. This single unit may be reexpressed in conventional units using the  joule as unit of energy or the kg as unit of mass, so
\vspace{-2mm}
that
\be
\label{um}
\hbox{\em unit of mass}\ \,u_m\,= \,\hbar/{c^2}\ \, \hbox{s}^{-1}\,=\, {\hbar_\circ}/c_\circ^2\ \, {\rm kg} \,=\, 1.173\;369\;392\;016\,... \times 10^{-51}  \ \hbox{kg}\,.
\ee

\noindent
The unit of angular momentum, now dimensionless, is 1. 
This may be verified from
$u_m=(\hbar/c^2)\ \hbox{s}^{-1},\  u_l= c \ {\rm s},\ u_t = 1 $\,s, combining
 into
{\em unit of angular momentum}\ \,$u_j= u_m\, u_l^2/u_t\,= \hbar\,$.
It may be reexpressed,
 in conventional units, as 
$
(h_\circ/2\pi) \ \hbox{J\,s}=1.054\;571\;817\;646\,... \times 10^{-34}$ J\,s\,.

\vspace{2.5mm}

For a theoretician, measuring space directly in seconds and masses and energies in s$^{-1}$ would be 
the natural thing to do.
This is not very practical however, the second as a unit of distance being too large, while the s$^{-1}$ as a unit of energy (close to $10^{-34}$ J) or  unit of mass (close to $10^{-51}$ kg)  is too small.
The unit of angular momentum when $\hbar=1$, now the dimensionless number 1, is suitable for individual photons, electrons or nucleons,
 but too small for macroscopic objects.

\vspace{2.5mm}

It is thus convenient to  resize consistently the above units (\ref{ul0},\ref{ue},\ref{um}) of length, energy, and mass to turn them into more convenient ones, still keeping  the familiar names of metre, joule, and kilogram in order not to disrupt long habits and use easily former measurements.
The quantities $c$ and $h$ or $\hbar$, often previously referred to as ``fundamental constants of nature'', get 
 fixed, according to the
 \vspace{-1mm}
  resolution \cite{bipm}
\be
\label{ch}
\ba{c}
\hbox{``\,\em the speed of light in vacuum $c$ is \,299\;792\;458  \,{\rm m/s}'',}
\vspace{.1mm}\\
\hbox{\em ``\,the Planck constant h is \,6.626\;070\;15 $\times$\,10$^{\,-\hbox{\scriptsize {34}}}$  {\rm J\,s}''.}
 \ea
 \ee
 \vspace{2mm}
 \noindent
 and used as normalization constants in the redefinition of the system of units.

 \vspace{1mm}
 
 Just as we can use the above choice of $c$ to redefine the metre from the natural unit of length $u_l$ as in (\ref{ul0}), we can use the choice of a fixed value of $h$ (or $\hbar$), namely $h=h_\circ$\,J\,s, to redefine the joule and the kilogram from the natural units and energy and mass in (\ref{ue},\ref{um}). This leads to
 \be
\label{ume2}
\left\{\ \ba{cclllcc}
1 \ \hbox{m}&=&\!\! 
1/c_\circ\ \ \,\hbox{\em unit of length} \ \ u_l\,,
\vspace{.5mm}\\
1\ \hbox{J}&= &\!\! 
1/\hbar_\circ \ \ \hbox{\em unit of energy}\ \ u_e\,, 
\vspace{.5mm}\\
1 \ \hbox{kg} &= &\!\!  
c_\circ^2/\hbar_\circ\ \,\hbox{\em unit of mass}\ \ u_m\,,  \ea \right.
\ee

\vspace{0mm}

\noindent
which define the metre, the joule, and the kilogram.

\section{\boldmath Fixing $\,h= 6.626\;070\;15\times 10^{-34}$ J\,{\lowercase {s}\,}, \ \u{AND}\ \boldmath {\normalsize $\,\hbar $ =\;1}}

 Beyond that, we can identify $u_l=u_t=1$\,s, 
 and $u_m=u_e=1$\,s$^{-1}$. This leads to {\em identify\,} the metre, the joule, the kilogram, and the newton, etc., as fixed numbers of \,s, s$^{-1}$ or   s$^{-2}$, etc..
Indeed, beyond the official definitions
$c= c_\circ\, \hbox{m\,s}^{-1}$ with $c_\circ\ \hbox{fixed at}  \ 299\;792\;458$,
$h=h_\circ\;  \hbox{J\,s}$  with $h_\circ\ \hbox{fixed at 6.626\;070\;15}  \times 10^{-34} $, we now
\vspace{-2mm}
identify
 \be
\label{ume02}
\left\{\
\ba{ccclcccccc}
c\!&=& c_\circ\ \hbox{m\,s}^{-1} \!& 
\equiv \ 1\ ,
\vspace{2mm}\\
\hbar \!&=&\!\!\hbar_\circ\  \hbox{J\,s}\!& 
\equiv\ 1\ ,
\ea\right.
\ \  \Longrightarrow\ \ 
\left\{\
\ba{ccllcccccc}
{\hbar}/{c^2}\!&=&\!\!{\hbar_\circ}/{c_\circ^2}\  \ \hbox{kg s}\!&& \equiv& \  1 \ ,
\vspace{1mm}\\
{\hbar}/{c}\!&=&\!{\hbar_\circ}/{c_\circ}\  \hbox{kg m}\,=\,{\hbar_\circ}/{c_\circ}\  \hbox{N\,s}^2\!\!\!
&&\equiv&\  1\  ,
\vspace{1mm}\\
\hbar {c}\!&=&\!\ \hbar_\circ{c_\circ}\  \ \hbox{J\,m}\!&& \equiv &\  1\  .
\ea\right.
\ee
This 
leads to the identifications
\be
\label{ume04}
\dis 1\ \hbox{metre}\,=\,\frac{1}{c_\circ}\ \hbox{s}\,,\ \ \ 1\ \hbox{joule}\,=\,\frac{1}{\hbar_\circ}\ \hbox{s}^{-1}\,,\ \ \ 1\ \hbox{kilogram}\,=\,\frac{c_\circ^2}{\hbar_\circ}\ \hbox{s}^{-1}\,, \ \ \ \ \hbox{1 newton}\,=\,\frac{c_\circ}{\hbar_\circ}\ \hbox{s}^{-2}\,,\
\ee
or, explicitly,
  \be
\label{ume03}
\framebox [15.9cm]{\rule[-1.8cm]{0cm}{3.9cm} $ \dis
\ \left\{ \! \ba{ccccl}
1 \ \hbox{m}\!\!&=&&&\ \ \ \ \  ({1}/{299\;792\;458})\ \ \hbox{s}\,,\ \ \ \ 
\vspace{2mm}\\
1\ \hbox{J}\!\!&=& \ \dis \frac{2\pi}{6.626\;070\;15\times 10^{-34}} \ \, \hbox{s}^{-1}
&=&\dis
 \ \,  .948\;252\;156\;246\,... \times 10^{34}\ \ \hbox{s}^{-1} ,
  \vspace{2mm}\\
  1 \ \hbox{kg} \!&=&\ \dis \frac{2\pi\,\times (299\;792\;458)^2}{6.626\;070\;15 \times  10^{-34}}   \ \,\hbox{s}^{-1} &=&\dis
 \ \,  .852\;246\;536\;175\,... \times 10^{51}\ \ \hbox{s}^{-1} ,
    \vspace{2mm}\\
\ 1\,{\rm N}=1\, {\rm kg\,m\,s}^{-2}= 1\, \hbox{J\,m}^{-1}\!&=&\dis \ \frac{2\pi\,\times 299\;792\;458}{6.626\;070\;15 \times  10^{-34}}   \ \,\hbox{s}^{-2} &=&\dis
  2.842\;788\;447\;250\,... \times 10^{42}\  \hbox{s}^{-2} .
\vspace{2mm}\\
\ea \right.
$}
\ee

\vspace{2mm}
\noindent
We can then say that 
the electron is a spin-1/2 particle, with $\hbar =\hbar_\circ$ J\,s = 1, the speed of light being
$c = $ 299\;792\;458 m/s, and, at the same time, $c=1\,$, with $\,\u{\mu_\circ}=\mu_\circ$~N/A$^2$ = 1.

\vspace{4mm}

\noindent
{\bf \em \boldmath Should we fix $h$, or rather  {\normalsize \boldmath $\hbar$}?}\ 

\vspace{2mm}

The above equations (\ref{ume2}-\ref{ume03}) also indicate that, at least from a conceptual point of view, {\em it could have  been preferable to fix
numerically  the value of $\,\hbar$}  (possibly to 1.054\;571\;82 $\times \,10^{-34}$ J\,s), which determines directly the angular momentum of the photon ($\hbar$), or of the electron ($\hbar/2$), rather than fixing $\,h$. This would also have led to simpler definitions 
 for the joule, the kilogram, and the newton in (\ref{ume2}-\ref{ume03}).

\vspace{2mm}

Anyhow, once such a choice has been made, the definitions of
the joule, and of the associated kilogram, can also be formulated through one of the sentences

\vspace{-2mm}

\be
\ba{c}
\left\{
\ba{lcc}
\hbox{\em the angular momentum of an electron is } \ \ 
\dis \frac{1}{2}\  \frac{6.626\;070\;15 \times 10^{-34}}{2\pi}\ \ \hbox{J\,s \ \ (or kg\,m}^2\,\hbox{s}^{-1})\,,
\vspace{3mm}\\
\hbox{\em the angular momentum of a circularly polarized photon  is } \ \dis \frac{6.626\;070\;15 \times 10^{-34}}{2\pi}\ \  \hbox{J\,s \  (or kg\,m}^2\,\hbox{s}^{-1})\,,
\ea \right.
\ea
\ee

\vspace{5mm}
\noindent
which give a direct physical meaning to the fixing of Planck's constant $h$ in (\ref{hc},\ref{ch}).

\vspace{2mm}

Still, defining in this way the metre, the joule, the kilogram, and the newton from the second, the s$^{-1}$, or the s$^{-2}$ also requires 
a {\it mise en pratique} allowing one to  apply these definitions to macroscopic objects.
The Kibble balance, in particular, appeals to electrical measurements 
relying on electromagnetic interactions in the quantum regime, implying the Josephson and  von Klitzing constants $K_J = 2e/h$ and $R_K= h/e^2$
\cite{ki}.
One can also count a larger number of atoms in a crystal, and determine the mass of a silicon sphere as $m= (8 \,V/a^3) \,m({\rm Si})$, where $m({\rm Si})$ is the mean mass of a silicon atom in the crystal, and $a^3$ the volume of the unit cell,  with eight atoms on average \cite{silic}.

\vspace{15mm}

\noindent
{\bf \em The redefinition of the kilogram \vspace{1.5mm} from quantum mechanics \hbox{does not rely on interactions}}

\vspace{1mm}

The definition of the kilogram from quantum mechanics through fixing the value of $h$ or $\hbar$, however,  
does not in principle appeal to interactions, and in particular does not require the explicit consideration of electromagnetic interactions. 
This is true even at the macroscopic level, as we shall see with the Casimir force, which depends only on $\hbar$ and $c$,  and whose expression is fixed if once $\hbar c$ is fixed.

\vspace{2mm}

Indeed the momentum of a particle, and thus its mass, or the energy of a photon, may be determined from the dimension associated with the diffraction or interference pattern associated with it.
This involves only free particles. 
Indeed the Dirac equation for a free spin-1/2 field or particle, the Klein-Gordon equation for a free spin-0 field or particle, and the Maxwell equations for a free massless spin-1 field, 
\vspace{-2mm}
expressed as 
\be
\left\{ \ \ \ba{ccc}
\hbox{Dirac\,:} &
\displaystyle  \ (\,i\,\hbar\ \gamma^\mu\,\partial_\mu-\,mc\,)\, \psi\,=\, 0\ ,&\ \ \ \ 
\vspace{1mm}\\ 
\hbox{Klein-Gordon\,:}  &\displaystyle   (\,\hbar^2\ \partial^\mu\partial_\mu+m^2c^2\,)\,\varphi\, =\, 0\ ,
&\ \ \ \ 
\vspace{1mm}\\ 
\hbox{Maxwell\,:} &
\partial_\mu F^{\mu\nu}=\,0\ ,& \ \ \ \ \ \ \  \
\ea \right. 
\ee
involve the constants $c$ and $\hbar$ associated with the definition of the system of units, but not the interactions. 
With $c=\hbar=1$ (a choice compatible with $c$ = 299\;792\;458 m/s and $h$ = 6.626\;070\;15 $\times 10^{-34}$ J\,s as we saw), these equations get simply written as
\be
\label{free}
 (\,i\,\partial\!\!\!/-m)\, \psi=0\,,\ \ \
 (\,\Box+m^2)\ \varphi=0\ ,\ \ \
\partial_\mu F^{\mu\nu}= 0\,.
\ee
 {\em These fundamental equations no longer involve the ``fundamental constants'' $c$ and $\hbar$!}  
 \ \,But their numerical values in SI units, $c_\circ$ and $\hbar_\circ$, now serve in the definition of the fundamental units of length and mass.
 
\vspace{2mm}

The consideration of the Casimir force between two conducting plates allows in principle for passing from the microscopic to the macroscopic level for the measurement of a force and thus also a mass, defined as earlier by fixing $h$ or $\hbar$. Ideally the Casimir force depends only on geometry, and associates to a distance $l$ between two conducting plates a force per unit surface, i.e.~a pressure. Its expression involves $c$ and $\hbar$.  Once $c$ is fixed it may be used, in principle, to determine the value of $\hbar$ in ordinary units of J\,s, even if only with a very modest precision. Or conversely, if $\hbar$ and thus $\hbar c$  is fixed, it allows to realize the joule from the metre.
Considered as a force per unit surface it reads 
\be
{\cal P}\,=\,\frac{F}{S}= \, -\,\displaystyle \frac{\pi^2}{240}\ \,\frac{\hbar c}{l^4}\,\simeq\,-\,\dis \frac{.013 \ \hbox{dyn/cm}^2}{l\,(\mu \hbox{m})^4} \,\simeq \,
\displaystyle \frac{1.3 \ \hbox{mN/m}^2}{l\,(\mu{\rm m})^4}\ .
\ee
It indicates how a very small force might be realized experimentally at the macroscopic level, even if not precisely, once
$\hbar$ is fixed and the unit of length has been defined by fixing $c$\,.
This allows us, in principle, to pass at the macroscopic level from a geometric unit of distance to a mechanical unit of force and, subsequently, energy and mass, without having to consider explicitly the value of the elementary charge $e$ (but taking into account  boundary conditions for the electromagnetic field between the plates).

\vspace{2mm}
In practice the precision of quantum electromagnetic effects is essential to get a precise determination of $h$ through $K_J=2e/h$ and $R_K=h/e^2$, and taking full advantage of fixing $h$ \cite{ki}. This motivates a more precise definition of electrical units than the usual one through the traditional definition of the ampere\,\cite{bipm,mise,poiri}.

\ghost{
\vspace{7mm}
\bc
*\hspace{8mm}*
\vspace{4mm}
*
\ec
}

\vspace{0.5mm}

\section{Electrical units as tied to the mechanical ones}

\subsection{The ampere and the other electrical units, as obtained from \boldmath $\mu_\circ$}

The electrical units are  at the moment  rigidly tied to the mechanical ones through a fixed $\mu_\circ$ as seen in Section \ref{sec:elec}.
Since 1946 and up to now in 2018, the ampere A has been defined as follows \cite{amp,amp2}:
\be
\label{A0}
\ba{c}
\hbox{\em ``\,The ampere is that constant current which, if maintained in two straight parallel conductors} 
\vspace{.1mm}\\
\hbox{\em of infinite length, of negligible circular cross-section, and placed  \,{\rm 1~m} apart in vacuum,}
\vspace{.1mm}\\
\hbox{\em would produce between these conductors a force equal to $2 \times 10^{-7}$ newton per metre of length.''}
\ea
\ee
\vspace{-1mm}

\noindent
This originates from Amp\`ere's force law stating that the force $F$ per length $L$ between two such parallel conductors at distance $r$, through which a current of intensity $I$ passes, is
$\,F/L=(\mu_\circ/{2\pi})\, I^2/{r}\,$.
This definition involves the fixed number $2\times 10^{-7}$ associated with the factor 
$\mu_\circ$, referred to as the magnetic permeability of free space,  conventionally taken as $4\pi\times 10^{-7}$ N/A$^2$ in the present SI (but soon to become $4\pi\times 10^{-7}\,\eta^2$ N/A$^2$).
\vspace{2mm}

Amp\`ere's force law, however, is a fundamental law of physics, and  it seems more logical to write it independently of the somewhat arbitrary parameter $\mu_\circ$, which determines the sizes chosen for the electrical units (very much as done for eqs.\,(\ref{free}) when writing the free equations of motions independently of $c$ and $\hbar$).
We shall thus discard the factor $\mu_\circ$ from its expression
 (or write it with an implicit parameter $\u{\mu_\circ}=1$),
 to include it
as a numerical dimensionless coefficient ($4\pi\times 10^{-7}$ in the present SI) 
 \vspace{-.2mm}
within the man-made definition of the ampere. This allows for a new equivalent formulation, in which 
$1\,{\rm A}^2\!=\mu_\circ\,\rm N $\,,
 the ampere being  obtained as 
proportional to a square root of the newton as in (\ref{mua}),
\be
\label{A1bis0}\
\hbox{\em Amp\`ere's force law}\ \ \ \  \frac{F}{L}\,=\,\frac{I^2}{2\pi\,r}\ \  \ \Longrightarrow\  \  \ \ 1\, \hbox{A}\, =\,\sqrt{\mu_\circ\, \hbox{N}}\ .  \hspace{1.5cm}
\ee
This provides
the force per unit length between two conductors carrying currents of 1\,A at a distance of 1 metre in vacuum,
obtained from (\ref{A1bis0}) as $\,F= 1\,{\rm A}^2/2\pi= (\mu_\circ/2\pi) \,  {\rm N}$, i.e. at the moment, before the redefinition of the SI, as $2\times 10^{-7}$ N as in (\ref{A0}).

\vspace{2mm}

Writing as above fundamental laws with $\underline{\mu_\circ}=1$,
the ampere, the coulomb, and the volt get given by the new expressions
\vspace{-6mm}

\be
\label{ac}
\left\{\ \ba{lcl}
1\  \hbox{ampere} &=&\,\sqrt{\mu_\circ\!\hbox{\phantom{\large  I}\!N}\,}\,,
\vspace{3mm}\\
1\  \hbox{coulomb}\ \ =\ \ \ 1 \,{\rm A\,s} &=&\,\sqrt{\mu_\circ\,\, \hbox{kg}\hbox{\phantom{\large l}}\!\hbox{m}\,}\,,
 \vspace{1.5mm}\\
1\ {\rm volt}\ =\,1 \, {\rm J/C}=\,1 \, {\rm W/A}=\,\dis \frac{1}{\sqrt{\mu_\circ}} \, W/\sqrt{\rm N} &=& \dis\,\sqrt{\,\frac{\hbox{J\,m\,s}^{-2}}{\mu_\circ }}\,.
\ea\right.
\ee
The coulomb, in particular, is proportional to {\em a geometric mean of the kilogram and the metre}. 

\vspace{2mm}

The definitions (\ref{ac}), and (\ref{tesla0}) below,  of the various electrical units follow,  in agreement with the 
\hbox{($\mu_\circ$-indepen\-dent)} expressions for 
\vspace{-.6mm}
electrostatic energy ($qV$) or electrical power ($UI$) for the volt; Laplace's magnetic force 
on a  wire
($\vec {dF}= I\,\vec{dl}\times \vec B\,$) for the tesla; energy stored in a coil or capacitor ($LI^2\!/2$ or  $CV^2\!/2$) for the henry and farad; and power dissipated into a resistance ($RI^2$) for the ohm.
A magnetic field also appears as proportional to the square root of the corresponding energy density 
(expressed as $B^2/2$, with $\underline{\mu_\circ}=1$).
An inductance is proportional to a length and a resistance to a velocity, etc., with
\be
\label{tesla0}
\left\{\ \ba{ccccccl}
\hbox{1 tesla}\!&=&\! 1\ {\rm N/(A\,m)}\!&=&\!  \dis \frac{1}{\sqrt {\mu_\circ}}\ \sqrt{\,\hbox{N/m}^2} &=&\dis \frac{1}{\sqrt {\mu_\circ}}\ \sqrt{\hbox{J/m}^3}\,,
\vspace{1.5mm}\\
1\  \hbox{henry}\!&=&\! 1 \ {\rm J/A}^2\!&=&\! \dis \frac{1}{\mu_\circ} \ \hbox{J/N}&=&\, \ \dis \frac{1}{\mu_\circ}\ \ \ \, \hbox{m}\,,
\vspace{1.5mm}\\
1 \hbox{ farad}\!&=&\! 1 \, {\rm C/V}=1 \, {\rm J/V}^2\ &&\! &=& \ \  \mu_\circ\ \ \, \hbox{s}^2/{\rm m}\,,
\vspace{2.5mm}\\
1\  \hbox{ohm}\!&=&\! 1 \,{\rm V/A} = 1 \, {\rm W/A}^2\!&=&\! \dis \frac{1}{\mu_\circ}\ \hbox{W/N} &=&\ \, \dis \frac{1}{\mu_\circ}\ \ \, \hbox{m/s}\,.
\ea\right.
\ee

\vspace{2mm}
\noindent
The resulting expressions of electrical units are given in Table~\ref{tab:unit}.

\vspace{2mm}
We note the usual relations
\be
\label{hf}
1\,\rm H \,=\ 1\,\Omega\, s\,,\ \ \ \,1\,F\,=\,1\,s/\Omega\,,
\ee
in agreement with the characteristic times $\tau= L/R$ of an inductance and $\tau=RC$ of a capacitor in association with a resistance (with $1\,\rm H \times 1\,F = 1 \,s^2$, as from 
the relation $LC\omega^2\!=1$ for an $LC$ oscillator). These definitions and relations are by construction compatible with the rescaling of $\mu_\circ$ by a factor $\eta^2$, acting on the various units as in (\ref{varelec}).

\subsection{\boldmath Fixing $\,c=c_\circ$ m/s, $\hbar=\hbar_\circ$ J\,s, \,then also $\,c=\hbar=1$}

We can then take advantage of fixing $c=c_\circ\,  \rm m/s =1,\ \hbar=\hbar_\circ \, J\,s=1$, to express the ampere, and all electrical units, in terms of the second, s$^{-1}$ or s$^{-2}$, or even as fixed numerical constants, as displayed in Table \ref{tab:unit} of Section \ref{sec:elec}.
An impedance, $Z=R+j(L\omega-1/C\omega)$, gets dimensionless, and may be expressed in ohms, a dimensionless unit, equal to $1/\mu_\circ c_\circ$, with the ``impedance of vacuum'' being now
$\,Z_\circ =\mu_\circ c_\circ\ \Omega = 1$.

\vspace{2mm}
To understand things better we decompose the fixing of $c$ and $\hbar $ in two steps, first by fixing their values in m/s and J\,s as in the new SI \cite{bipm}, then by choosing also $c=\hbar =1$ as proposed here.
With 
$\,c=c_\circ$ m/s,  $\,\hbar=\hbar_\circ $ J\,s, we  express the ampere and coulomb as
  \be
  \label{achc}
  \left\{\ 
  \ba{ccccccccc}
  1\,\rm A \!&=&\! \rm \sqrt{\mu_\circ\,N} \!&=&\!\dis \sqrt{\,\frac{\mu_\circ\,\rm J\,s}{\rm m/s}}\ \ \rm s^{-1} \!&=&\!
  \dis \sqrt{\,\frac{\mu_\circ\,c_\circ}{\hbar_\circ}}\  \sqrt{\,\frac{\hbar}{c}}\ \ \rm s^{-1}\,,
  \vspace{2mm}\\
  1\,\rm C \!&=&\!\rm  \sqrt{\mu_\circ\,kg\,m} \!&=&\!\dis \sqrt{\,\frac{\mu_\circ\,\rm J\,s}{\rm m/s}} \!&=&\! \dis \sqrt{\,\frac{\mu_\circ\,c_\circ}{\hbar_\circ}}\ \sqrt{\,\frac{\hbar}{c}}\ .
  \ea\right.
  \ee
  Electric charges, including the coulomb,  still have a dimension, and may be evaluated at this stage in  $\sqrt{\,\rm kg\,m}$, with $\mu_\circ$ taken as dimensionless. 
  
  \vspace{2mm}
  
 With the additional choice   $c=\hbar=1$, the scale for all electrical units  gets fixed (in terms of $\mu_\circ$) and formulas (\ref{achc}) simplify into 
    \be
    \label{acac}
  \left\{\ 
  \ba{ccccccl}
  1\,\rm A \!&=&\! \rm \sqrt{\mu_\circ\,N}  \!&=&
  \dis \sqrt{\,\frac{\mu_\circ\,c_\circ}{\hbar_\circ}}\  \rm s^{-1}& =& \,1.890\,... \times 10^{18}\  \rm s^{-1}\,,
  \vspace{2mm}\\
  1\,\rm C \!&=&\!\rm  \sqrt{\mu_\circ\,kg\,m} \!&=&\! \dis \sqrt{\,\frac{\mu_\circ\,c_\circ}{\hbar_\circ}}
  \,=\,\sqrt{\frac{1}{\epsilon_\circ\,c_\circ\hbar_\circ}}& =& \,1.890\,... \times 10^{18}\,,
  \ea\right.
  \ee
  the coulomb becoming a dimensionless unit.
  Electrical charges are now dimensionless, as for angles, their natural unit being 1 in a system with $\hbar=c=\u{\mu_\circ}=\u{\epsilon_\circ}=1$. The coulomb is an artificial unit, resulting from the definition (\ref{A0}) of the ampere (from $\mu_\circ$), and of the metre and joule through the choices of $c_\circ$ and $\hbar_\circ$.
The elementary charge $e$, now dimensionless,  is expressed (as seen from (\ref{alphazero})) as
  \be
  \label{ealphazero}
  \framebox [12.5cm]{\rule[-.28cm]{0cm}{.8cm} $ \dis
  e\,=\,e_\circ \, {\rm C}\,=\,\sqrt{{e_\circ^2}/{\epsilon_\circ c_\circ\hbar_\circ}}\,=\,\sqrt {4\pi\alpha}\,
  = \,\sqrt {4\pi\alpha_\circ}\,\ \eta\,=\,.302\;822\;120\;789\,...\ \times \eta\ ,
  $}
  \ee
 thanks to the choice $\hbar=c  = \u{\mu_\circ}=\u{\epsilon_\circ} =1$.
 \vspace{-.5mm}
 The numerical value of $e$ is thus rescaled by the factor $\eta =\sqrt{\alpha/\alpha_\circ}$, very close to 1.

  \vspace{2mm}
  
At the same time, with $1\,\Omega = \rm 1\,W/A^2= (1/\mu_\circ) \, m/s=1/\mu_\circ c_\circ=\epsilon_\circ c_\circ$, the impedance of vacuum is
\be
\label{z00}
\framebox [8.5cm]{\rule[-.25cm]{0cm}{.75cm} $ \dis
\ Z_\circ=\mu_\circ c_\circ\ \Omega\,=\,376.\,730\;313\;461\,... \ \times \eta^2\  \Omega \,=1\ ,
$}
\ee
as it should with $c=\u{\mu_\circ}=\u{\epsilon_\circ}=1$\,.
\vspace{-.2mm}
 (The above numerical value in ohms has been evaluated with $\mu_\circ=4\pi\times 10^{-7}$, to be rescaled by $\eta^2$, with $Z_\circ$ remaining equal to 1.)\,
The impedance being now a dimensionless quantity its natural unit, 1,  is simply what is also referred to as the ``impedance of vacuum''.

\vspace{2mm}

From (\ref{tesla0}) and using 
$c= c_\circ $ m/s = 1, we find, with 1\,F = $\mu_\circ c_\circ^2 \rm \ m = (1/\epsilon_\circ) \,m$, the now symmetric expressions for the farad and the henry, 
given by
\be
\label{hf0}
\framebox [4.2cm]{\rule[-.25cm]{0cm}{.7cm} $ \dis
\rm \mu_\circ \,H\,=\,\epsilon_\circ \,F\,=\,1\ m\ ,
$}
\ee
or equivalently
\be
\label{hf1}
\mu_\circ c_\circ\,{\rm H}\,=\,\epsilon_\circ c_\circ \,\rm F\,=\,1\ s\ .
\ee

\pagebreak

\noindent
With $\,1\,\Omega= 1/\mu_\circ c_\circ=\epsilon_\circ c_\circ$ we recover eqs.\,(\ref{hf}), 
\be
\rm 1\,H/\Omega \,=\,1\,F\,\Omega\,=\,1\ s\ .
\ee

\noindent
For the tesla and V/m, we get from (\ref{ac},\ref{tesla0}) the analogous expressions
  \be
\label{vm}
\framebox [10cm]{\rule[-.9cm]{0cm}{2.1cm} $ \dis
\ \left\{\ 
\ba{ccc}
\rm 1\,T\!&=&\!
\hspace{2.0cm}
\,\dis \frac{1}{\sqrt{\mu_\circ }}\ \sqrt{\hbox{J/m}^3}\ ,
\vspace{.5mm}\\
\rm 1\,V/m\!&=&\!
\,\dis \frac{1}{\sqrt{\mu_\circ }}\  \sqrt{\hbox{J/m}^3}\ {\rm m/s}\,=\ \sqrt{\epsilon_\circ}\ \sqrt{\hbox{J/m}^3}\,=\,\frac{1}{c_\circ}\ \rm T\  ,
\ea\right.
$}
\ee 
thanks in the last case to the identification 1\,m = $(1/c_\circ)$ s.
This will be interpreted in subsection \ref{subsec:ener} in terms of the electric and magnetic field energy densities, 
now simply given by $E^2/2$ and $B^2/2$, corresponding to
 their standard SI expressions as $\,{\cal E}= \epsilon_\circ E^2/2$ and $B^2/2\mu_\circ$, respectively.

\subsection{A direct characterization of the coulomb}

Coulomb's law is now written with $\u{\epsilon_\circ}=1$ as $F\!=q^2/4\pi r$. 
\vspace{-.5mm}
With
$1\, \hbox{A}=\sqrt{\mu_\circ\, \hbox{N}}$ so that $1\, {\rm C}= \sqrt{\mu_\circ c_\circ^2 \, \hbox{N}}\ \rm m $ $= \sqrt{\,\hbox{\rm N}/\epsilon_\circ}\ \rm m$, one has
$1\,\rm C^2=1\,N\,m^2/\epsilon_\circ$.
The force between two charges of 1\,C at 1 metre apart in vacuum is recovered as $(1/4\pi\epsilon_\circ)$\,N.
We can duplicate  (\ref{A1bis0}) for the ampere, writing the parallel expression for the coulomb,

\vspace{-7mm}

\be
\label{claw}
\hbox{\em Coulomb's law}\ \ \  F\,=\,\frac{q^2}{4\pi r^2}\ \  \ \Longrightarrow\  \  \ \ 1\, \hbox{C}\, =\,\sqrt{\,\frac{\hbox{N}}{\epsilon_\circ}}\ \rm m\ .
\ee

\vspace{1mm}

This provides a physical understanding for the coulomb.
The electrostatic force 
between two charges $q=1$\,C, at $r =$ 1\,m apart in vacuum, is obtained as
$(1/4\pi\epsilon_\circ) \,{\rm N} = (\mu_\circ c_\circ^2/4\pi)\,   \hbox{N} = 10^{-7}\,c_\circ^2\, \eta^2\ {\rm N} =
 \,8.987\; 551\;787\,... \times 10^9$  N.
Their electrostatic energy $\,V\!= \mu_\circ c^2\,q^2/4\pi r$ is
\be
E\,=\,\frac{\mu_\circ}{4\pi}\ c_\circ^2\ {\rm J} =10^{-7}\  \eta^2\,(299\;792\;458)^2\ {\rm J}\,= \ 8.987\;551\;787\,\,... \times 10^9\ \eta^2\ {\rm J} \,.
\ee
This energy is no longer exactly known, owing to the new definition of the coulomb.
The equivalent mass,  according to relativity, is
\be
\frac{E}{c^2} = \frac{1}{c^2}\,\frac{q^2}{4\pi\epsilon_\circ  r}\,=\,\frac{\mu_\circ}{4\pi}\,\frac{q^2}{r}\,=\,10^{-7}\, \eta^2\ \frac{q^2}{r}\,.
 \ee
The coulomb may be characterized as follows:
\be
\ba{c}
``\, \hbox{\em The coulomb is such that the electrostatic energy} \\
\hbox{\em of two charges of \,$1{\rm \,C}$, $1{\rm \,m}$ apart in vacuum, is equal to $(1/4\pi\epsilon_\circ)$~{\rm J}.''}
\ea
\ee
\vspace{-6mm}
or:
\be
\label{C0}
\ba{c}
``\,\hbox{\em The coulomb is such that the mass equivalent to the electrostatic energy}
\\
\hbox{\em of two charges of  
\,$1{\rm \,C}$, $1{\rm \,m}$ apart in vacuum, is equal to $\mu_\circ/4\pi$ \,(\;{\em i.e.} $10^{-7}\,\eta^2$) {\rm kg}.''}
\ea
\ee
The factor $\eta$, at the moment equal to 1 within the present SI, takes into account  that 
the numerical value of the elementary charge $e$ is getting redefined as in (\ref{ee0}), with $\mu_\circ/4\pi$ very slightly shifted away from $10^{-7}$, proportionally to $\eta^2$.

\vspace{.5mm}

\section{Properties of the new system}

\subsection{\boldmath Invariance of the fine structure constant $\alpha$ under a rescaling of the fundamental units }

If we reintroduce explicitly the constant $\u{\mu_\circ}=\mu_\circ\ \rm N/A^2$ (momentarily considered as non-necessarily equal to 1), the ampere and coulomb become expressed as
\be
1\, \hbox{A}\, =\,\sqrt{\,(\mu_\circ/\u{\mu_\circ})\ \hbox{N}}\ , \ \ \ 1\, \hbox{C}\, =\,\sqrt{\,(\mu_\circ/\u{\mu_\circ})\ 
\hbox{kg\,m}}\ .
\ee
With $\,\hbar=\hbar_\circ$ J\,s,\ $c=c_\circ$\ m/s, so that ${\hbar}/{c}\,=\,(\hbar_\circ/{c_\circ})$ kg\,m,
we get, for the elementary charge,
\be
e^2\,=\,e_\circ^2\ {\rm C}^2\,=\,\frac{\mu_\circ e_\circ^2}{\u{\mu_\circ}}\ \frac{\hbar/c}{\hbar_\circ/c_\circ}\ ,
\ \ \ \ \ \hbox{so that}\ \ \ \ \ \
\frac{ \u{\mu_\circ} c \,e^2}{\hbar}\,=\,\frac{{\mu_\circ} c_\circ \,e_\circ^2}{\hbar_\circ}\,,
\ee

\vspace{-2mm}

\noindent
or

\vspace{-6mm}

\be
\alpha\,=\,\frac{e^2}{4\pi\u{\epsilon_\circ}\, \hbar c}\,=\,\frac{e_\circ^2}{4\pi\epsilon_\circ\hbar_\circ c_\circ}\,.
\ee
This provides an identity between the two expressions of the fine structure constant, associated, for any set of values of 
$\hbar,\,c,\,\u{\mu_\circ}$ (and corresponding $\u{\epsilon _\circ}$), with the rescaling of the fundamental units through the choice
$c=c_\circ\ \rm m/s,\ \hbar=\hbar_\circ\ J\,s, \ \u{\mu_\circ}=\mu_\circ\ N/A^2$.

\vspace{2mm}

If as a further step we choose $\hbar=c=\u{\mu_\circ}=\u{\epsilon_\circ}=1$,  we recover
\be
\alpha\,=\,\frac{e^2}{4\pi}\,=\,\frac{e_\circ^2}{4\pi\epsilon_\circ\hbar_\circ c_\circ}\,
\ee
leading to $\,e=\sqrt{4\pi\alpha}=.3028\,...\,$ as in (\ref{ealphazero}).

\subsection{\boldmath Invariance of the impedance of vacuum $Z_\circ$ under a rescaling of the fundamental units}
\label{sub:imp}

\label{subsec:ener}

Let us return to the SI for a short moment.
The energy density associated with an electromagnetic field,

\vspace{-5mm}

\be
\label{de0}
{\cal E}\,=\,\frac{\epsilon_\circ E^2}{2}+\frac{B^2}{2\mu_\circ}\,=\,
\,\frac{E^2+c^2 B^2}{2\mu_\circ c^2}\,=
 \sqrt{\,\frac{\epsilon_\circ}{\mu_\circ}}
\ \left(\,\frac{ E^2}{2c}+\frac{c B^2}{2}
\right)\ . 
\ee
with $\epsilon_\circ\mu_\circ c^2=1$, originates from the Lagrangian density $\,{\cal L }= - \,(1/4\mu_\circ) \,F_{\mu\nu}F^{\mu\nu}$.
The quantity
\be
Z_\circ\,=\,\sqrt{\,\frac{\mu_\circ}{\epsilon_\circ}}\,= \,\mu_\circ c
\ee
equal to the ratio $E/(B/\mu_\circ)$ for a plane electromagnetic wave in vacuum,
\vspace{-.5mm}
 is referred to as the impedance of vacuum. 
 \vspace{-.5mm}
 It appears as a normalisation coefficient in Mawxell's equations  $\vec{\hbox{rot}}\,\vec B\!= \mu_\circ c\ (\vec j/c)\,+ ...\,$, \linebreak $\hbox{div} \,\vec E/c = \rho/\epsilon_\circ c$, which fix the relative scale of $(\vec E/c, \,\vec B)$ compared to that of the current density $(\rho, \,\vec j/c)$, source of the electromagnetic field. 
 \vspace{-.3mm}
 This may also be seen from Poisson's equation for a static field, expressed as $\Delta (V/c,\,\vec A)= -\,\mu_\circ c \ (\rho,\,\vec j /c)$, with \,div\,$\vec A= 0$.

\vspace{2mm}

With $\mu_\circ$ expressed in N/A$^2$ or H/m and $c$ in m/s, $Z_\circ$  is obtained in the SI in $\Omega$. We have at present
\be
Z_\circ\,= \,\mu_\circ c
\,=\,(4\pi\times 10^{-7} \ {\rm H/m}) \,(c_\circ\, {\rm m/s}) =4\pi\times 10^{-7} \,c_\circ\ \Omega\,=\,376.\,730\,...\ \Omega\,.
\ee
But if the elementary charge gets fixed as in (\ref{ee0}),  $\mu_\circ$ gets changed into $4\pi\times 10^{-7} \,\eta^2$.
The measure in ohms of the impedance of vacuum is multiplied by $\eta^2$, to become in the new SI
\be
Z_\circ\,= \,4\pi\times 10^{-7} \,\eta^2\,c_\circ\ \Omega=\,376.\,730\,... \ \eta^2\ \Omega\,.
\ee
But this does not mean that the impedance of the vacuum has been multiplied by $\eta^2$! At the same time, with $1\,\Omega$ = 1\,W/A$^2$, the size of the ohm is divided by $\eta^2$ as in (\ref{varelec}), according to 
\be
\Omega\,=\,\Omega_{\rm old}\,/\eta^2\ ,
\ee
so that $Z_\circ$ itself, equal to $\mu_\circ c_\circ\,\Omega$,  remains unchanged.  As for the elementary charge $e$, expressed either as $e_{\rm old}\,\rm C_{old}$ or as $e_\circ\,\rm C$, $Z_\circ$, which is an intrinsic quantity, may be expressed equally well as 
\be
Z_\circ\,= \,\underbrace{4\pi\times 10^{-7} \,c_\circ}_{\hbox{\normalsize \ 376.\,730\,...}}\ \Omega_{\rm old}\,= \, 4\pi\times 10^{-7} \,\eta^2
\,c_\circ\ \Omega\,.
\ee

\vspace{1mm}

This ensures that the  impedance of vacuum  $Z_\circ$ (but not its measure in ohms!), is indeed insensitive to the change in electrical units following from the fixing of $e=e_\circ$\,C within the new SI, and remains insensitive to the value of the elementary charge $e$ \,-- i.e. to the value of $\alpha$ to be measured experimentally in the future.
This is fortunate, as $Z_\circ$ is an intrinsic quantity that should not depend on the chosen system of units, nor on future measurements of $\alpha$.
 But it also illustrates how fixing the numerical value of the electrical charge in the new SI as in (\ref{ee0}), while practically convenient, makes life more complicated as far as explaining what the impedance of the vacuum really is, and for which reason its measure should now depend on $\alpha$.

\vspace{2mm}

In the new proposed system  their is no such concern, as $1\,\Omega = 1/\mu_\circ c_\circ$ and $\,Z_\circ=\mu_\circ c_\circ\,\Omega = 1$, independently of any rescaling of $\mu_\circ$ in agreement with (\ref{varelec}). This is a consequence of 
$\,\hbar=c=\u{\mu_\circ}=\u{\epsilon_\circ}=1$ so that the impedance of vacuum is the unit of impedance, namely simply 1.
 The Lagrangian and energy densities
 for the free electromagnetic field get simply expressed as
 \be
 \label{lag}
 {\cal L }\,=\, - \,\frac{1}{4} \ F_{\mu\nu}F^{\mu\nu},\ \ \ \ \ \ {\cal E}\,=\,\frac{E^2+B^2}{2}\ ,
 \ee
as usual in relativistic quantum field theory, and its equation of motion reads $\,\partial_\nu F^{\mu\nu} \!= j^\nu$.

\subsection{The volt/m, the tesla and the electromagnetic energy}

Expressions   (\ref{vm}) of the tesla and  volt/metre read
\be
\label{tv}
{1\ \rm T  \,=\,\dis \frac{1}{\sqrt{\mu_\circ}}\ \,\sqrt{\,\hbox{J/m}^3}\,,
\ \ \ \ \ 
1\,V/m\,=\,\dis \sqrt{\epsilon_\circ}\ \sqrt{\,\hbox{J/m}^3}}\,=\,\frac{1}{c_\circ}\ \rm T\,.
\ee
We reobtain correctly the energy densities for unit magnetic and electric fields of 1\,T and 1\,V/m,  
obtained from (\ref{lag}) as
\vspace{-2mm}
\be
\label{edensb}
{\cal E} \,=\,\frac{B^2}{2} = \,\frac{ 1\ {\rm T}^2}{2}= \,\frac{1}{2\mu_\circ} \  {\rm J/m}^3\,,\ \ 
\ \ \hbox{and}\ \ \ \ {\cal E} \,=\,\frac{E^2}{2} = \,\frac{ 1\ {\rm (V/m)}^2}{2}=\,\frac{\epsilon_\circ}{2}
\  {\rm J/m}^3\,.
\ee

\vspace{1mm}
\noindent
This also  provides a simple characterization of the tesla and volt/metre, 
\vspace{2mm}as follows:
\be
\label{entesla}
\left\{\ 
\ba{c}
``\,\hbox{\em The tesla is the magnetic field corresponding to an energy density in vacuum}
\vspace{.1mm}\\
1/2\mu_\circ \ ({\rm or}\  10^7/8\pi\eta^2)\ \ \hbox{J/m$^3$.\,''}
\vspace{1.5mm}\\
``\,\hbox{\em The volt/metre is the electric field corresponding to an energy density in vacuum}
\vspace{.1mm}\\
\epsilon_\circ/2 \ ({\rm or}\  10^7/8\pi c_\circ^2\eta^2)\ \ \hbox{J/m$^3$.\,''}
 \ea \right.
 \ee

 \vspace{2mm}

\noindent
The energy density for a  magnetic field of of 1\,T is larger than for an electric field of 1\,V/m by a large factor $1/\mu_\circ\epsilon_\circ =\,c_\circ^2\simeq \,9 \times 10^{16}$, as easily understood since 
1\,T = $c_\circ$\,V/m.

\vspace{2mm}
Let us also consider the Poynting vector, now simply expressed as $\vec P= \vec E \times \vec B\,$.
\,For orthogonal unit electric and magnetic fields of 1\,V/m and 1\,T, the energy flux density per unit time would be
\be
\label{flux}
{\rm 1\,V/m \times 1\,T\,} =\,\sqrt{{c_\circ}/{\mu_\circ \hbar_\circ}} \ \,{\rm s^{-2}}\times \sqrt{{c_\circ^3}/{\mu_\circ \hbar_\circ}}\ \,{\rm s^{-2}}
 \,=\,\frac{1}{\mu_\circ}\ \rm W/m^2 \,=\,\frac{1}{4\pi\times10^{-7}\ \eta^2}\ \rm W/m^2\,,
\ee
with $\,c_\circ^2\ \rm s^{-2}\!= 1\,m^{-2}$, and $1/\hbar_\circ$ = 1\,J\,s.
 This is evaluated in a more conventional way in the SI with $\,\vec P= \vec E \times \vec B/\mu_\circ$, as
\be
\frac{\rm 1\,V/m \times 1\,T}{\mu_\circ} \,=\,\rm \frac{1\ (W/A\,m) \ (N/A\,m)}{4\pi\times 10^{-7}\ \eta^2\ N/A^2}\,=\,\frac{1}{4\pi\times 10^{-7}\ \eta^2}\ \rm W/m^2\,.
\ee
Eq.\,(\ref{flux}) illustrates how the factor $4\pi\times 10^{-7}\ \eta^2$, instead of being present in the expression of the Poynting vector as in the standard formulation,
 gets now included within the new expressions of the  electrical units as  in (\ref{tv}).

\vspace{2mm}

Electrical units thus remain strongly tied to mechanical ones and completely fixed by them, up to the scale factor $\,\mu_\circ =4\pi\times 10^{-7} \, \eta^2$, where $\eta$ is very close to 1.
These formulas reflect the tight connection between electrical and mechanical units, originating from the traditional definition of the ampere. This connection gets somewhat weakened as $\mu_\circ$ is no longer fixed if the value of the elementary charge gets fixed as in (\ref{ee0}). 
\vspace{-.5mm}
Electrical units are then no longer  rigidly tied to mechanical ones, but in a more flexible way, 
proportionally to $\sqrt{\mu_\circ}\,$ \,i.e. to $\eta = \sqrt{ \alpha/\alpha_\circ}\, $ for the ampere and the coulomb, as described by eqs.\,(\ref{varelec}).

\subsection{Inductances and capacitances}

The henry may be defined from the magnetic energy stored in a coil, $E=LI^2/2$, so
that
\be
1\,{\rm H}\,=\,1\,{\rm J/A}^2\,=\,\frac{1}{\mu_\circ}\  {\rm J/N \,=\,\frac{1}{\mu_\circ} \ m}\,=\,\frac{1}{\mu_\circ c_\circ} \ \,{\rm s\,=\frac{1}{(376.\,730\,...\,\eta^2)}\ \,s}\,.
\ee
The inductance $L$ for a long coil of length $l$, inner area $S$ and $N$ turns is now simply given, 
\vspace{-.5mm}
with $\u{\mu_\circ}=1$, 
by $L= N^2S/l\,$.
For a coil of unit length $l$ = 1\,m the inductance 
per unit inner core area $S$ (supposed small)
\vspace{-2mm}
 and unit $N^2$ is
\be
L\,=\, 1\,{\rm m}\,=\,\mu_\circ\, \rm H\,,
\ee
as found with the usual SI formula $L=\mu_\circ\,N^2\,S/l\,$.
Similarly, the capacitance of a plane capacitor with empty space between the plates is given, with $\u{\epsilon_\circ}=1$, by $C=S/l$. 
 For a unit distance between the plates $l$ = 1\,m, the capacitance per unit surface $S$ is 
\be
C\,=\,1\,{\rm m}\,= 
\,\epsilon_\circ\,{\rm F}\,,
\ee
as seen from (\ref{hf0}), 
and as  found with the usual SI formula $C=\epsilon_\circ\,S/l\,$.
The impedance  $\sqrt{L/C}\,$ evaluated with the above values 
$L=1\,\rm m=\mu_\circ \,H$ and $C=1 \,\rm m= \epsilon_\circ \,F$ is the impedance of free space, 
which is 1:
\be
\sqrt{{L}/{C}}
\,=\,\sqrt{\,{\mu_\circ \rm H}/{\epsilon_\circ \,\rm F}}\,=\,\mu_\circ c_\circ \ \sqrt{\rm {H}/{F}}\,=\,\mu_\circ c_\circ \ \Omega\,=\,Z_\circ\,=\,1\ .
\ee

\subsection{The weber and the quantum of flux}

The weber is 
the magnetic flux induced by a current of 1\,A circulating in a coil of inductance $L = 1$\,H,
\be\;
{\rm 1 \ Wb \,=\, 1\,H \times 1 \,A} \,=\,{1}/{\sqrt{\mu_\circ c_\circ \hbar_\circ}}\,= \,\rm 1\,V\,s\,= \,5.\,017\;029\;284\;119\,...  \times 10^{15}\ \eta^{-1}\,.
\ee
It is a dimensionless unit, as for the coulomb and the ohm, equal to 1\,V\,s (as \,1\,H\,=\,1\,$\Omega$\,s).
These three dimensionless units,
\be
\left\{\ \ba{ccccc}
1\,\rm C\!&=&\!\sqrt{{\mu_\circ c_\circ}/{\hbar_\circ}}\!&=&\!1.890\,... \times 10^{18}\ \eta\,,\
\vspace{2mm}\\
1\,\Omega\!&=&\!1/\mu_\circ c_\circ \!&=&\!1/(376.7\,...\,\eta^2)\,,
\vspace{2mm}\\
\rm 1\,Wb&=&1/\sqrt{{\mu_\circ c_\circ}{\hbar_\circ}}&=& \,5.017\,...\times 10^{15}\ \eta^{-1},
\ea\right.
\ee
satisfy the relation, independent of $c_\circ,\,\mu_\circ$ and $\hbar_\circ$, 
\be
1\,\rm CÊ\times 1\,\Omega\,=\,1\,Wb\,,
\ee
reflecting that $1\,\rm AÊ\times 1\,\Omega\,=\,1\,V$, or 1\,F\,$\times \,1\,\Omega$ = 1\,s\,.

\vspace{3mm}

The magnetic flux quantum, now exactly fixed in webers, is
thus
\be
\Phi_\circ= \frac{h}{2e}  \,=\,\frac{\pi\hbar_\circ}{e_\circ} \ \underbrace{{\rm J\,s/C}}_{\hbox{\small Wb}}\,=\,\pi\,
\sqrt{\frac{\hbar_\circ}{\mu_\circ c_\circ e_\circ^2}}  
\,=\,\sqrt{\frac{\pi}{4\alpha}}\,=\,\frac{\pi}{e} \,\simeq\,10.\,374\;382\;97\, ,     
\ee 
\vspace{-4mm}
or 
\be
\Phi_\circ= \,h/2e\,=\,2.\,067\;833\;848\;461\,... \times 10^{-15} \ {\rm Wb}\,=\frac{\pi}{e}\,\simeq\,10.\,374\;382\;97\ .
\ee
We simply recover  $\pi /e$ with $e=\sqrt {4\pi\alpha}\,$, at no surprise since $\hbar =c=\u{\mu_\circ}=\u{\epsilon_\circ}=1$. 
The inverse of the flux quantum is the Josephson constant,
\be
K_J\,=\,\frac{1}{\Phi_\circ}\,=\,\dis \dis\frac{2e}{h} \,=\,\dis \dis\frac{2e_\circ}{h_\circ}\  \, {\rm Wb}^{-1}\,=\,483\;597.\,848\;416\,...  \ \ {\rm GHz/V} \,=\, \dis\frac{e}{\pi} \,\simeq \,.0963\;912\;748\ .
\ee
It is related to the expression of \,1\,eV = $\pi\ (2e_\circ/ h_\circ) \ \rm s^{-1} = 1.519\,... \times 10^{15}\ s^{-1}$, \,as we shall see in (\ref{Jev0}-\ref{Jev}).

\vspace{2mm}
The expression of the physical laws, now written without reference to $\hbar,\,c,\,\mu_\circ$ and $\epsilon_\circ$, and the new expressions for the corresponding units, are given in Table \ref{tab:form}.

\begin{table}[t]
\caption{\ The physical laws written in a universal way, with no reference to $\hbar, \, c, \, {\mu_\circ}$ and ${\epsilon_\circ}$.  This includes Maxwell's equations. $\hbar_\circ,\, c_\circ,\, \mu_\circ$ and $\epsilon_\circ$ define our usual units by scaling appropriately the natural ones obtained from the second. 
The usual formulas including $\mu_\circ$ and $\epsilon_\circ$ are shown in the last column. Due to the fixing of $e=e_\circ $\,C,  $\,\mu_\circ$ can no longer be fixed at $4\pi\times 10^{-7}$ but should be multiplied by 
$\eta=\sqrt{\alpha/\alpha_\circ}$, which is very close to 1 (cf. eqs.\,(\ref{alphac}-\ref{etaz})).
\vspace{5mm}
\label{tab:form}}
\begin{tabular}{c}
$\ba{|c|ccccccccccc|}
\hline
&&&&&&&&&&&\vspace{-3mm} \\
\ba{c}\vspace{-3mm}\\ 
\hbox{Physical law}\\ \vspace{2.5mm}\hbox{or expression}\ea  &&&& \hbox{\hspace{-8cm}New expression\hspace{-8cm}}   &&&&&&&\!\!\hbox{\hspace{-1.2cm}Usual expression\ \ \ } 
\\  [3mm]
\hline\hline
&&&&&&&&&&&\\  [-1mm]
\hbox{Amp\`ere's force law}  &\ \  \dis \frac{F}{L}\!\!\!\!\!\!&=&\!\dis \frac{I^2}{2\pi r} \   & \hbox{with} &  1\,{\rm A} \!&=&\! \sqrt{\,\mu_\circ \phantom{\hbox{\normalsize I}} \!\rm N} & \ \ \longleftrightarrow \  \ & \dis \frac{F}{L}\!&=&\!\dis \frac{\mu_\circ\,I^2}{2\pi r} \ \ \ \ \ \ \ \ \ \ \ \ \ \
\\ [4mm]
\hbox{Coulomb's law}  &\ \ F \!\!\! \!\!\!&=&\!\ \dis \frac{q^2}{4\pi r^2} \  &\ \hbox{''} &
1\,\rm C \!&=&\!\dis \sqrt{\,\frac{\rm N}{\epsilon_\circ}}\ \rm m & \longleftrightarrow  & F\!\!&=&\!\dis \frac{q^2}{4\pi\epsilon_\circ r^2} \ \ \ \ \ \ \ \ \ \ \ \ \
\\ [4mm] \hline
&&&&&&&&&&&\\  [-3mm]
\hbox{electric energy density}  & \ \  {\cal E} \!\!\!\!\!\!&=&\! \dis \frac{E^2}{2}\    &\ \hbox{''} &
1\,\rm V/m \!&=&\! \dis \sqrt{\epsilon_\circ\, \rm J/m^3} & \longleftrightarrow  & {\cal E}\!\!&=&\! \dis \frac{\epsilon_\circ\,E^2}{2}\ \ \ \ \ \ \ \ \ \ \ \ \
\\ [3mm]
\hbox{\ magnetic energy density\ }  &\ \   {\cal E}\!\!\!\! \!\!&=&\! \dis \frac{B^2}{2}\   &\ \hbox{''} &
1\,\rm T \!&=&\!  \dis \sqrt{\,\frac{\rm J/m^3}{\mu_\circ}} & \longleftrightarrow  & {\cal E}\!\!&=&\! \dis \frac{B^2}{2\mu_\circ}
\ \ \ \ \ \ \ \ \ \ \ \ \ \
\\ [4mm]
\hbox{\ Poynting vector\ }  &\ \  \vec {\cal P}\!\!\!\!\!\!&=&\! \vec E  \times \vec B&\ \hbox{''} &
\!1\,\rm (V/m)\, T \!\!&=&\!  \dis \frac{\rm W/m^2}{\mu_\circ} & \longleftrightarrow  & \vec {\cal P}\!\!&=&\! \dis
\frac{\vec E \times \vec B}{\mu_\circ}\ \ \ \ \ \ \ \ \ \ \ \ \
 \\ [4mm]
\hbox{\ Lagrangian density\ }  &\ \   {\cal L}\!\!\!\!\!&\!\!\!\!\!\!=&\! \!\!\! \dis -\,\frac{1}{4}\,F_{\mu\nu} F^{\mu\nu} \!\!\!\! \!\!\!\!\!\!\!\!\!&\hbox{} &
 \!\!& &\!   & \longleftrightarrow  &  {\cal L}\!\!&=&\! \dis
\frac{\epsilon_\circ\,E^2}{2}-\,\frac{B^2}{2\mu_\circ} \ \ \ \ \ 
\\ [4mm] \hline
&&&&&&&&&&&\\  [-2mm]
\ \hbox{cap. of plane capacitor} \  &\ \ C \!\!\!\!\!\!&=&\!\ \dis \frac{S}{l} &\ \hbox{''} &
1\,\rm m \!&=&\!\epsilon_\circ \rm \,F & \longleftrightarrow  & C\!\!&=&\!  \dis \frac{\epsilon_\circ\, S}{l} \ \ \ \ \ \ \ \ \ \ \ \ \ \ 
\\ [4mm]
\hbox{inductance of coil}  &\ \ L \!\!\!\!\!\!&=&\!\ \dis \frac{N^2S}{l} &\ \hbox{''} &
1\,\rm m \!&=&\!\mu_\circ \rm  \,H  & \longleftrightarrow  & L\!\!&=&\! \dis \frac{\mu_\circ \,N^2 S}{l} \ \ \ \ \ \ \ \ \ \ \ \ \ 
\\ [4mm] \hline
&&&&&&&&&&&\\  [-3mm]
\hbox{impedance of vacuum} & \ \ Z_\circ\!\!\!\!\!\!&=&\!\ 1  &  && & & \longleftrightarrow  & Z_\circ\!\!&=&\! \! \dis \sqrt{\,\frac{\mu_\circ}{\epsilon_\circ}}\ \, \Omega\simeq \,377\ \Omega \
\\ [4mm] 
\hbox{fine structure constant} & \, \dis\alpha=\frac{e^2}{4\pi}\!\!\!&=&\!\dis \frac{e_\circ^2}{4\pi}\ \rm C^2 \!\!\!&\ \hbox{''}  & 1\,\rm C \!&=&\! \dis\sqrt{\,\frac{1}{\epsilon_\circ c_\circ\hbar_\circ}}& \longleftrightarrow  & \alpha\!\!&=&\!  \dis \frac{e_\circ^2}{4\pi\epsilon_\circ\hbar_\circ c_\circ}\simeq\frac{1}{137} \ 
\\ [5mm] \hline
\ea$
\end{tabular}
\vspace{4mm}
\end{table}

\subsection{Relating energies with distances and  times}

The spectrum of the hydrogen atom, for example, may be obtained
with $\hbar=c=\u{\epsilon_\circ}=1$ and $e=\sqrt{4\pi\alpha}$, which leads directly to the Rydberg energy 
\be
1\, {\rm Ry}\,=\,\frac{1}{2}\, m_e \, ({e^2}/{4\pi})^2\,=\,\frac{1}{2}\, m_e \, \alpha^2\,\simeq\,13.\,605\;6930\ \rm eV\,.
\ee
We can use the fine structure constant  $\alpha=e^2/4\pi\simeq 1/137.\,035\;9991\,...\,$, and express the mass of the electron as $m_e\simeq $\ .510\;998\;946\ MeV,  with no need to refer to the velocity of light.

\vspace{2mm}

A useful formula to relate energies and distances is obtained by evaluating the product
1\,eV $\times$ 1\,m,
\vspace{-4mm}
 from
\be
1\,=\,(\hbar_\circ\,{\rm J\,s})\,(c_\circ\,{ \rm m/s})\,=\,
\frac{\hbar_\circ c_\circ}{e_\circ}
\ {\rm eV\,m}\,=\,10^9\ \frac{\hbar_\circ c_\circ}{e_\circ}\ {\rm MeV\  fm}\,,
\ee
using
\be
10^9\ \frac{(6.\,626\;070\,15\times 10^{-34})Ê\times (2.\,997\;924\;58 \times 10^8)}{2\pi\times 1.\,602\;176\;634 \times 10^{-19}}\,=\,197.\,326\;980\;459\;302\,...\ \ \ \ \hbox{(now exactly known)\,.}
\ee
\vspace{-2mm}

\noindent
This may be remembered, not surprisingly, as
197.\,327\ {\rm MeV\ Fermi} $\simeq $ 1\,,
or 
\be
\label{mevfm}
1\ \rm MeV^{-1}=\, 197.\,326\;980\,459\,...\ fm\,.
\ee
To give some illustrative examples, the Lyman $\alpha$ wavelength, Bohr radius, reduced Compton wavelength of the electron, range of weak interactions and Planck length may  be expressed directly using (\ref{mevfm}), with no reference to $\hbar$ and $c$ factors, as
\be
\label{dist}
\left\{\ \
\ba{lcccccc}
\hbox{Lyman $\alpha$ wavelength}\hspace{-30mm}&& &&\!\dis
\hbox{\Large $\left(\right.$}  \!\approx\, \dis \frac{4}{3}\ (1+\frac{m_e}{m_p})\ 
\frac{2\pi}{13.606\ \rm eV}\hbox{\Large $\left.\right)$}\!&\simeq &\dis 1215.67\ \hbox{\AA}\ ,
\vspace{2mm}\\
r_B\!&=&\!\dis \frac{1}{m_e\alpha}\!&\simeq &\!\dis\frac{137.\,035\;999}{.510\;998\;946\ \rm MeV}\!&\simeq &\dis.529\;177\;21\ \hbox{\AA}\ ,
\vspace{2mm}\\
\lambda_e\!\!\!\!\! / \! \ \ \ &=&\!\dis \frac{1}{m_e}\!&\simeq &\!\dis\frac{1}{.510\;998\;946\ \rm MeV}\!&\simeq &\dis \ \ \ 3.\,861\,5927 \times 10^{-13}\ \rm m\ ,
\vspace{2mm}\\
\lambda_W\!\!\!\!\!\!\!\!/ \! \ \ \ &=&\!\dis \frac{1}{m_W}\!&\simeq &\!\dis\frac{1}{80.38\ \rm GeV}\!&\simeq &\dis \ \ \ 2.\,455\times 10^{-18}\ \rm m\ ,
\vspace{2mm}\\
l_P\!&= &\!\dis\dis \frac{1}{m_P}= \sqrt{G_N}\,&\simeq &\!\dis\frac{1}{1.\,2209\times 10^{19}\ \rm GeV}\!&\simeq &\dis  \ \ \ 1.\,6162\times 10^{-35}\ \rm m\ .
\ea\right.
\ee

\vspace{4mm}

The relation between energy and time comes from 1\,J\,=\,$(1/\hbar_\circ)\, {\rm s}^{-1} = .948\,252\,156\,... \times 10^{34}\  \rm s^{-1}$, 
\vspace{-3mm}
so that
  \be
  \label{Jev0}
 1\,{\rm eV}\,=\,e_\circ \,{\rm J} \,=\,\frac{e_\circ}{\hbar_\circ}\ \rm s^{-1}\,=\,1.519\;267\;447\;878\;626\,... \times  10^{15}\  s^{-1}=1/(6.582\,119\,569\;509\;065\,...
  \times 10^{-16}\  s)\,.
 \ee
also equivalent to 
 \be
1\ \rm MeV^{-1}=\, 197.\,326\;980\,459\,302\,...\ fm\,=\,6.582\,119\,569\;509\,... \times 10^{-22}\,{\rm s}\,.
\ee
\vspace{-1mm}

1\,eV\,s/$\pi=2 e_\circ/h_\circ= .483\,597\,848\,416\,983\,... \times  10^{15}$ involves the same factor
as for $K_J=(2e_\circ/h_\circ)$ Hz/V.
We have the equivalence between the new expression of $K_J$ as equal to $e/\pi$, and the well-known expression 
of $\hbar$, precisely evaluated as $(\hbar_\circ/e_\circ)\ \rm eV \,s=6.582\;119\;569\;509\,... \times 10^{-22}\ \rm MeV\,s$, and now equal to $\hbar =1$:

\vspace{-2mm}
\be
\label{Jev}
\framebox [14.4cm]{\rule[-.55cm]{0cm}{1.3cm} $ \dis
\ba{ccc}
K_J=\,.483\,597\,... \times 10^{15}{\ \rm Hz/V}\,=\,e/\pi\,
\vspace{1mm}\\
\hspace{-2mm}\Longleftrightarrow \hspace{4mm} 1\ \rm eV = \pi \times .483\;597\;848\;416\,... \times 10^{15} \ s^{-1}=\,1.519\;267\;447\;878\,...\times 10^{15} \  s^{-1}\ .
\ea
$}
\ee

\vspace{0mm}

\subsection{The kelvin and the Boltzmann constant}

In the new SI, the unit of thermodynamic temperature, the  kelvin K, will be derived from the unit of energy by fixing the numerical value $k_\circ$ of the Boltzmann constant $k$ \cite{bipm}, so that
\be
\label{k}
k =1.\,380\;649 \times 10^{-23} \ \rm J/K\,.
\ee
We can then go one step further. 
Very much as space and time are related and may both be measured in seconds, or as energy may be measured in s$^{-1}$ when time is measured in seconds, thermodynamic temperature and energy are related and may both be measured with the same unit, the s$^{-1}$, by choosing a unit value of the Boltzmann constant $k\,$. 
 The Maxwell-Bolzmann distribution, for example, will then be simply expressed  with $k=1$, proportionally to $e^{-E/T}$.
\vspace{2mm}

Choosing $k=1$  is compatible with fixing its value in SI units according to (\ref{k}), leading to
\be
k = 1.\,380\;649 \times 10^{-23}\ \rm J/K \,=\,1\,.
\ee
This allows to identify  the kelvin K with a certain number $k_\circ$ of joules according to
\be
\framebox [5.2cm]{\rule[-.25cm]{0cm}{.7cm} $ \dis
\rm \,1\,K \,=\,1.\,380\;649 \times 10^{-23} \ J\,.
$}
\ee
We can also express energies in eV, rather than in joules. We then have the exact  relation
\be
\rm 1\,eV= 1.\,602\;176\;634 \times 10^{-19} \ \rm J\,=\, \frac{1.\,602\;176\;634 \times 10^{-19}}{1.\,380\;649 \times 10^{-23}}\ K
\,=\, 11\,604.\,518\,121\, ... \ K\ ,
\ee
remembered as 1\,eV $\simeq$ 11\,605 K.

\subsection{A more symmetric treatment between electricity and magnetism}

In this discussion, and within the SI, $\mu_\circ$ has been given a favored treatment  as compared to $\epsilon_\circ$, owing to the original definition of the ampere.
But there is no special reason to do so,
\vspace{-.8mm}
 as they contribute in  similar ways 
to the expressions of the speed of light $c= 1/\sqrt{\,\u{\epsilon_\circ}\phantom{l}\u{\mu_\circ}} \,$ and  impedance of \vspace{-.8mm}
vacuum $Z_\circ= \sqrt{\,{\u{\mu_\circ}}\ba{c} \vspace{-6mm}\\  \!/ \!\vspace{-0.2mm}\\ \ea {\u{\epsilon_\circ}}}\ $.
To get a system with $c=\u{\mu_\circ}=\u{\epsilon_\circ}=1$, as proposed here, we may use, equivalently,
 two of the four sets of conditions
\be
\label{zc}
\left\{\ \ba{cclcccc}
c \!&=& c_\circ \rm \ m/s\!&=&\!1\ ,
\vspace{1mm}\\ 
Z_\circ   \!&=&  z_\circ \rm \ \Omega \!&=&\!1\ ,
\vspace{1mm}\\
\u{\mu_\circ}  \!&=&\mu_\circ \rm \ H/m = \mu_\circ c_\circ \rm \ H/s \!&=&\!1\ ,
\vspace{1mm}\\
\u{\epsilon_\circ}   \!&=& \epsilon_\circ \rm \ \, F/m \,= \,\epsilon_\circ c_\circ  \ \,\rm F/s\!&=&\!1\ ,
\ea \right.
\ee

\vspace{1.5mm}

\noindent
where the last three are associated with eqs.\,(\ref{z00}-\ref{hf1}).
For example we already have the equivalences
\be
\left\{\ \ba{ccccccccccc}
\vspace{-5.5mm}\\
c \!&=&\! c_\circ \rm \ m/s\ ,
\vspace{1mm}\\ 
Z_\circ   \!&=&\!  z_\circ \rm \ \Omega \ ,
\ea \right.
\ \ \Longleftrightarrow \ \ \
\left\{\ \ba{ccccccccc}
\u{\mu_\circ}  \!&=&\!  \dis  \frac{Z_\circ}{c}\!&=&\! \dis \frac{z_\circ}{c_\circ}\ \ \frac{\Omega \,\rm s}{\rm m}\!&=&\!\mu_\circ\ \rm H/m\!&=&\!\mu_\circ c_\circ  \ \rm H/s\ ,
\vspace{1mm}\\
\u{\epsilon_\circ}   \!&=&\!  \dis  \frac{1}{c\,Z_\circ}\!&=&\! \dis \frac{1}{c_\circ z_\circ}\ \ \frac{\rm s}{\Omega\,\rm m}\!&=&\!
\! \epsilon_\circ \rm \ F/m\!&=&\!\epsilon_\circ c_\circ  \ \rm F/s\ ,
\ea \right.
\ee
\vspace{1mm}

\noindent
using relations (\ref{hf}) between units,  1\,$\Omega$\,s = 1\,H, \ 1\,s/$\Omega$ = 1\,F, together with
\be
\left\{\  \ba{ccc} c_\circ \!&=&\! \dis {1}/{\sqrt{\,\mu_\circ \epsilon_\circ}}\ ,
\vspace{2mm}\\
z_\circ \!&=&\! \dis \sqrt{\,{\mu_\circ}/{\epsilon_\circ}}\ ,
\ea\right.
\ \ \Longleftrightarrow\ \ \ 
\left\{\  \ba{ccc} \mu_\circ \!&=&\! \dis {z_\circ}/{c_\circ}\ ,
\vspace{2mm}\\
\epsilon_\circ \!&=&\! \dis  {1}/{c_\circ z_\circ}\ .
\ea\right.
\ee
  
This leads to the equivalence between the two sets of conditions in (\ref{zc}),  where electricity and magnetism play similar roles,
\be
\label{czmu}
\left\{\ \ba{ccccccc}
c \!&=&\! c_\circ \rm \ m/s\!&=&\!1\ ,
\vspace{1mm}\\ 
Z_\circ   \!&=&\!  z_\circ \rm \ \Omega \!&=&\!1\ ,
\ea \right.
\ \ \Longleftrightarrow \ \ \
\left\{\ \ba{ccccccc}
\u{\mu_\circ}  \!&=&\! \mu_\circ c_\circ \rm \ H/s\!&=&\!1\ ,
\vspace{1mm}\\
\u{\epsilon_\circ}   \!&=&\! \epsilon_\circ  c_\circ \rm \ F/s\!&=&\!1\ .
\ea \right.
\ee
The fixing of $\u{\mu_\circ}$ or $\u{\epsilon_\circ}$ to 1 may also be viewed, equivalently, as a fixing of the impedance of vacuum to 1, next to $c=1$.
\vspace{2mm}

We have  formulated this  analysis by imposing $c=1$ and $\u{\mu_\circ}=1$.
But we could also have selected, equivalently, any two of the above sets of conditions.
With 1\,F/m  = 1\,C$^2$/J\hspace{.18mm}m, the supplementary set of equations in (\ref{czmu}) may be used (optionally) as follows:
  \be
 \left\{ \  \ba{ccccccccccc}
  Z_\circ  \!&=&\!z_\circ \rm \ \Omega \!&=&\! 1\,,  &\ \hbox{to fix } &\ \ 1\,\Omega\!&=&\! \dis {1}/{z_\circ}\,=\,{1}/{\mu_\circ c_\circ}\,=\,\epsilon_\circ c_\circ\ ,
  \vspace{2.5mm}\\
  \u{\epsilon_\circ}   \!&=&\! \epsilon_\circ \  \dis {\rm C^2}/{\rm J\hspace{.3mm}m}\!&=&\!1\,,  & \ \hbox{to fix }  
  &\ \ 1\,\rm C\!&=&\! \dis \sqrt{{\rm \,J\,m}/{\epsilon_\circ}}\,=\,{1}/\sqrt{\epsilon_\circ \hbar_\circ c_\circ}\ ,
  \ea
  \right.
  \ee
  
  \vspace{2mm}
  \noindent
  with the same results as from the fixings of $c$ and/or $\u{\mu_\circ}$.
  
  \vspace{2mm}

 However, due to the fixing of the elementary charge as $e=e_\circ\,$C, the quantity $z_\circ= \mu_\circ c_\circ =1/\epsilon_\circ c_\circ$ must be kept ``floating'' to adapt to the new value of the ohm, just as for $\mu_\circ$ to adapt to the size of the ampere, and $\epsilon_\circ$ to that of the coulomb,
 with $z_\circ\propto \eta^2$, $\mu_\circ\propto \eta^2$ and $\epsilon_\circ \propto \eta^{-2}$ as in (\ref{varelec}).

\subsection{\boldmath The classical limit}

The fundamental laws of nature may now be written in a universal way, no longer referring to the parameters $ \hbar,\,c,\, \u{\mu_\circ}\,,\,\u{\epsilon_\circ}\,,\,k, ...$, which can all be taken equal to 1, nor to the specific numerical values $ \hbar_\circ,\,c_\circ,\, {\mu_\circ},\,{\epsilon_\circ},\,k_\circ, ...$ which appear in their expressions when SI units are used.
Maxwell's equations, in particular, no longer involve $\mu_\circ$ and $\epsilon_\circ$, now included within the definitions of the electrical units. The Lorentz transformations relating space and time (or magnetism and electricity) within the framework of relativity can also be written without reference to the speed of light $c$, now equal to 1, through the relation $c=c_\circ $ m/s = 1. The numerical parameter $c_\circ$ gets included within the definition of the metre as obtained from the second, providing the same results as with the usual formalism, thanks to the relation 1\,metre = $(1/c_\circ)$ second.

\vspace{2mm}

In such a framework with $c=\hbar=1$  Einstein's formula $E=mc^2$ for a free particle at rest further simplifies into $E=m$. For a particle of mass $m$ and momentum $p=k$ we have
\be
\label{en}
E\,=\,\sqrt{m^2+p^2}\,=\,\sqrt{m^2+k^2} \,=\,\omega\ .
\ee
Now that $c=1$ one may enquire about ``taking the classical limit" to recover a non-relativistic situation.
This can no longer be done by taking a limit ``$c\to \infty$", but is now obtained in the limit of small velocities as compared to 1. For small $v=d\omega/dk= dE/dp=p/E\,$, \,eq.\,(\ref{en}) provides back the non-relativistic expansion of the mechanical energy, as $E=m+ p^2/2m\,+\,...\ $, without having to consider a limit in which the speed of light would become very large.
This is similar for quantum effects. 
We can no longer consider a limit for which ``$\hbar\to 0$''. Still, with  $\hbar=1$ as the quantum of action (or angular momentum), the classical limit now corresponds to situations involving large values of the action,
as compared to the unit quantum.

\vspace{-1mm}

  \subsection{The mole and the Avogadro constant}

Let us now mention the SI unit of ``amount of substance'' or mole.
In the new SI the mole gets defined by fixing the Avogadro constant to $N_A= 6.022\;140\;76 \times 10^{23} \rm \ mol^{-1}$ \cite{bipm}. 
If $n$ is the amount of substance (in moles) in a sample of X, the number of elementary entities 
 is $N=n\,N_A$.

\vspace{2mm}

But the most natural unit for counting ``entities'' is 1, which fits well in a system of units with  $\hbar=c=\u{\mu_\circ}=\u{\epsilon_\circ}=Z_\circ =k=1$.  We shall then fix the Avogadro constant at ${N_A}=1$, as we did for 
 $\hbar, \, c,\ \u{\mu_\circ},\ \u{\epsilon_\circ}, \ Z_\circ$ and $k\,$ by fixing their numerical values in SI units to be
 $\hbar_\circ,\,c,\,{\mu_\circ},\,{\epsilon_\circ},\,z_\circ $ and $k_\circ$, respectively.
The Avogadro constant $N_A$ may then be fixed at  $N_{A\circ} \rm \ mol^{-1}$ as in the SI, and, at the same time at $N_A=1$.
\vspace{-4mm} 
This implies
\be
N_A \,=\,\underbrace{6.022\;140\;76 \times 10^{23}}_{\hbox{$N_{A\circ}$}} \rm \ mol^{-1}\,=\,1\ .
\ee

\vspace{-1mm}

\noindent
This leads us to identify the mole with a pure number, namely the Avogadro number,
\be
\framebox [6.5cm]{\rule[-.25cm]{0cm}{.7cm} $ \dis
1\,{\rm mol}\,=\,N_{A\circ}\,=\,6.022\;140\;76 \times 10^{23}\,.
$}
\ee
The number of elementary entities of a substance X in a sample of $n$ moles is simply $N=n \,N_{A\circ}$.

\vspace{2mm}
 The Avogadro constant $N_A$, normally expressed in mol$^{-1}$, no longer appears as a fundamental constant of nature, but rather as a counting device. Fixing it to the natural unit, 1, as for the other constants, makes the mole appear as a fixed number, like saying that there are twelve eggs in a dozen of eggs. But it is a very large number, the Avogadro number, fixed for consistency with past definitions to  $N_{A\circ}= \,6.022\;140\;76 \times 10^{23}$.

\vspace{2mm}

We may also consider the candela, SI unit of luminous intensity in a given direction, even if we cannot really view it as a fundamental unit. It is now defined and normalized by taking the luminous efficacy of a monochromatic radiation of frequency $540 \times 10^{12}$ Hz
(and wavelength $\lambda \simeq .555\  \mu$) to be  \linebreak 
$K\rm_{cd}=683\ cd\,sr \,W^{-1}$ \cite{SI,bipm}.
Fixing it at 1 as for the other constants, we get 
$
K\rm_{cd}=683\ lm \, W^{-1}\!=683\ cd\,sr \,W^{-1}\!=1\,.
$
The candela and the lumen appear, for the specific radiation considered, as a certain number of watts per steradian or watts, with 1\,cd = ({1}/{683})\ W/sr\,, 1\,lm = (1/683)\ W.

\vspace{4mm}

\section{Conclusions}

  \vspace{2mm}

The International System of units is getting redefined, with 
the joule and kilogram obtained by fixing the Planck constant at
$h=6.626\;070\;15 \times 10^{-34}$ \rm J\,s\,.
This makes obsolete the international prototype of the kilogram stored at BIPM. No longer having to rely on such a single material object  for the definitions of mechanical and electrical units is a huge progress.

\vspace{2mm}
The coulomb and the other electrical units are also redefined, so that the elementary charge is fixed at
$e = 1.602\;176\;634 \times 10^{-19}\ \rm C$
in the new SI. 
This  requires that the vacuum magnetic permeability $\mu_\circ$  be 
\linebreak slightly adapted, into  $\mu_\circ =4\pi\times 10^{-7}\, \eta^2$ N/A$^2$.
Fixing the values of $h$ and $e$ in joule$\,\cdot\,$seconds and coulombs will allow for more precise measurements, thanks to quantum electrical metrology based on the Josephson and quantum Hall effects.
Still one may regret that, with $\mu_\circ$ no longer exactly fixed but proportional to $\alpha$, all electrical units, together with
the numerical values of the vacuum magnetic permeability (in N/A$^2$ or H/m), electric permittivity (in F/m) and vacuum impedance
(in ohms) become dependent on $\alpha$.

\vspace{2mm}

An appealing system should have $c=\u{\mu_\circ}=\u{\epsilon_\circ}=1$ and
$\hbar=1$, as suggested by relativity and quantum mechanics;
but this is usually considered as unpractical 
as it would naturally lead to units of space and mass $\,\simeq 3\times 10^8$~m and $ 10^{-51}$ kg, not very convenient.
Still
it is possible to reconcile and unite both systems,
allying the practical interest and convenience of the normal SI units to the advantages and elegance of a symmetric system with 
$\,c=\u{\mu_\circ}=\u{\epsilon_\circ}=\hbar=1\,$. 
 \vspace{2mm}
 
 To this end the fundamental laws of physics, universal, may be expressed without referring to $c$ and $\hbar$, nor to the convention-dependent parameters  $c_\circ,\, \hbar_\circ,\, \mu_\circ$ and $\epsilon_\circ$.
These now serve to define and resize appropriately our fundamental units of length, energy and mass, and intensity, all defined from the second or s$^{-1}$,  so that we recover our usual units, metre, joule and kilogram, ampere, suitably normalized.

\vspace{2mm}
This is done, first, through a rewriting of the  laws of electromagnetism by eliminating  $\mu_\circ$, and subsequently $\epsilon_\circ$,
from their expressions. 
$\mu_\circ$  becomes  a conventional dimensionless normalisation coefficient for the electrical units, getting included within their expressions.  The ampere and the coulomb get given
 by
\be
\label{ackgm}
1\,{\rm A}\,=\,  \sqrt{\,\mu_\circ \phantom{\hbox{I}} \!\rm N}\,,\ \ \ 1\,\rm C
\,=\,   \sqrt{\,\mu_\circ\phantom{\hbox{I}} \!{\rm kg\,m}}  \ .
\ee
This is sufficient to derive all electrical units from a choice of $\mu_\circ$. This one
 is initially fixed to $4\pi\times 10^{-7}$,  multiplied by $\eta^2$, very close to 1,    
to allow for the necessary adjustement of the ampere and the coulomb,
if the elementary charge gets fixed as in (\ref{ee0}). This new formulation is well adapted to take in charge that
 the ampere and the coulomb are no longer rigidly tied to the newton, but  allowed to  slightly ``float'' proportionally to $\sqrt{\mu_\circ}$\,, \,i.e. to 
 $\eta =\sqrt{\alpha/\alpha_\circ}\,$.

\vspace{3mm}

By demanding that $c$ and $\hbar$, in addition to being numerically fixed in SI units
according to the official definitions,
be also equal to 1,
we can unite the advantages of both systems  through the equations
\be
\label{hcmu10}
\left\{\ \ba{cccccccccl}
c\!&=&\! c_\circ \ {\rm m/s}\!\!&=&\! 1&&\ \ \Rightarrow\ \ & 1\, \rm m\!&=&\ \ \ \  ({1}/299\;792\;458) \ \rm s\,,
\ \ \ \ \ 
\vspace{1.3mm}\\
\hbar\!&=&\! \dis \hbar_\circ \ \, {\rm J\,s}\!\!&=&\! 1&&   \Rightarrow& 1\  \rm J &=&
\ \ \ \  ({1}/{\hbar_\circ}) \ \rm s^{-1} =\, .948\,252\,... \times 10^{34}\ \rm s^{-1}\,,\ \ \ 
\vspace{1.3mm}\\
\u{\mu_\circ}\!&=&\!\! \mu_\circ \ {\rm N/A}^2\!\!&=&\! 1&&   \Rightarrow&\rm 1\,A&=& \!\!\sqrt{\,\mu_\circ \phantom{\hbox{\large I}} \!\!\rm N}\,= \dis \sqrt{{\mu_\circ c_\circ}/{ \hbar_\circ} }\ \ {\rm s}^{-1}=\,1.\,890\,067\,... \times 10^{18}\ \rm s^{-1}\,.
\ea\right.
\ee
One can also, equivalently, define directly the dimensionless ohm and coulomb through one of the conditions  \be
 \left\{\ \ba{ccccccccccl}
  Z_\circ  \!&=&\!z_\circ \rm \ \Omega\! \!&=&\! 1  & \ \   \Rightarrow &\ \ 1\,\Omega\!&=&\!  {1}/{z_\circ}\,=\,{1}/{\mu_\circ c_\circ}\,=\,\epsilon_\circ c_\circ \!&=&\!  1/376.\,730\,...\ ,
  \vspace{2mm}\\
  \u{\epsilon_\circ}   \!&=&\! \epsilon_\circ \  {\rm C^2}/{\rm N\,m^2}\!\!&=&\!1  &  \ \   \Rightarrow
  &\ \ 1\,\rm C\!&=&\! \dis \sqrt{{\rm N\,m^2}/{\epsilon_\circ}}\,=\,\sqrt{{1}/{\epsilon_\circ \hbar_\circ c_\circ}}
   \!&=&\! 1.\,890\,067\,... \times 10^{18}\,,
  \ea
  \right.  \hspace{8.8mm}
  \ee
allowing for a more symmetric treatment between electricity and magnetism.

\vspace{2mm}

The coulomb, the ohm, and the weber, related by 1\,C $\times \,1\,\Omega$ = 1\,Wb, become dimensionless, with the elementary charge $e$, the impedance of vacuum $Z_\circ$, and the flux quantum $\Phi_\circ $ expressed as pure numbers:
\be
\left\{
\ba{cclcc}
e&=&\ \ \  \,1.\,\,602\;176\;634 \times 10^{-19}   \ {\rm C}&=&\sqrt{4\pi\alpha}\ \simeq\ .302\;822\;1208\ ,
\vspace{1.5mm}\\
Z_\circ&=&\mu_\circ c_\circ \ \Omega\,= \,376.\,730\;313\ ...\ \Omega\ \ \ \ \ \  &=&\ \ \ \ \ \ 1\ ,
\vspace{1.5mm}\\
\Phi_\circ &=& \dis {h}/{2e} =  \ 2.\,067\;833\;848\,... \times 10^{-15}\  {\rm Wb}    &=&
\ \ \dis{\pi}/{e} \ \ \,\simeq \,10.\,374\;382\;97 \ .
\ea\right.
\ee

\vspace{2.5mm}

The impedance of vacuum is 1, as it should with $\u{\epsilon_\circ} =\u{\mu_\circ} =1$.
  \vspace{-.1mm}
  This is an improvement over the new SI description in which $Z_\circ =\mu_\circ c_\circ \ \Omega$  refers to an ohm dependent on $\mu_\circ$ and thus now on $\alpha$, hiding that $Z_\circ=1$ is the natural unit of impedance.
  Fixing the impedance of the vacuum to $Z_\circ\! = \,376.\,730\ ... \ \Omega = 1$ also 
determines the von Klitzing constant, both in ohms and in terms of the fine structure constant, as $R_K=h/e^2=1/2\alpha\,$. Its SI expres\-sion in ohms is simply recovered as 376.\,730\,313\,...\,$\Omega$ (vacuum impedance)\,$\times \,137.\,035\;9991\,...\,/2\,$, i.e.
\be
R_K = h/e^2=  376.\,730\,313\,... \ \Omega/2\alpha = 25\,812.\,807\,459\, ... \ \Omega\,=\,1/2\alpha\,\simeq\, 68.\,517\;999\;57\, .
\ee
The Josephson constant, now also dimensionless, is
\be
K_J = 1/\Phi_\circ = 2e/h = 483\;597.\,848\,416\, ... \  {\rm GHz/V} = e/\pi\,\simeq\, .0963\;912\;748\,.
\ee
Inductances and capacitances are naturally expressed in metres owing to their geometric origin.
They can be converted in ordinary SI units, with inductances measured in henrys proportionally to $\mu_\circ$, and capacitances in farads proportionally to $\epsilon_\circ$, thanks to the 
symmetric relations

\vspace{-4mm}
\be
\rm 1\,m=\mu_\circ\, H= \epsilon_\circ\,F\,, \ \ \ \hbox{or}\ \ \ \rm 1\,s= \mu_\circ c_\circ \, H= \epsilon_\circ
c_\circ \,F\,.
\ee 
These provide back the equalities
$
\rm 1\,s= 1\,H/\Omega = 1\,F\,\Omega$\,, and \,1\,Wb = 1\,C$\times 1\,\Omega\,$ relating the dimensionless weber, coulomb, and ohm.
\vspace{-.3mm}
The volt/metre and tesla, both square roots of energy densities, proportional to $\sqrt{\rm J/m^3}$,   \;are obtained in s$^{-2}$.

\vspace{4mm}

Extending these ideas to the kelvin and the mole, we get a unified system embedding the new SI within a framework where quantities previously considered as ``fundamental constants of nature'' return to their natural status of being simply 1, i.e.
\be
\framebox [7.3cm]{\rule[-.25cm]{0cm}{.7cm} $ \dis
c\,=\,\hbar\,=\, \u{\mu_\circ}=\u{\epsilon_\circ}=Z_\circ \,=\,k\,=\,N_A\,=\,1\,,
$}
\ee
and no longer appear within the expressions of the fundamental laws.
This is achieved through the set of equalities, 

\vspace{-4mm}

\be
\label{111}
\framebox [9.4cm]{\rule[-.25cm]{0cm}{.7cm} $ \dis
\ c_\circ\, {\rm m/s \,=\,\hbar_\circ \, J\,s}\,=\,z_\circ\, \Omega\,=\,k_\circ\,{\rm J/K}\,=\,N_{A\circ} \ {\rm mol}^{-1}\,=\,1\ .
$}
\ee
 Once the second is chosen as the unit of time, $c_\circ$ fixes the size of the metre, $\hbar_\circ$ the joule and kilo\-gram, $k_\circ$ the kelvin, and $N_{A\circ}$ the mole.
$Z_\circ=1$ (or $\,\u{\mu_\circ}\,$ or $\,\u{\epsilon_\circ}=1$)   fix the vacuum impedance, magnetic permeability and electric permittivity to 1, and the size of the electrical units from $\mu_\circ$ (or the exact size of $\mu_\circ$ if the coulomb is obtained from the fixing of $e$).
\vspace{2mm}

We have chosen in (\ref{111})  to determine the ohm, a dimensionless unit, to be $1/z_\circ=1/\mu_\circ c_\circ
=1/376.730\,... $  so that the impedance of vacuum is $\,376.\,730\,... \ \Omega = 1$.
This also determines $R_K = 376.\,730\,... \ \Omega\,/2\alpha = 25\,812.\,807\, ... \ \Omega$.
We could choose as well, to fix the electrical units,  one of the equivalent conditions
\be
\label{112}
z_\circ \ {\rm H/s}\,=\,z_\circ^{-1} \ \rm F/s \,=\,\mu_\circ\ N/A^2 \,=\,\epsilon_\circ\  C^2/(J\,s\!\times \!m/s)\,=\,1\ .
\ee

\vspace{1mm}

It is often said that the new SI provides a system of units by fixing the values of fundamental constants of nature such as the speed of light and the Planck constant. Including it within the proposed system leads to transcend this point of view.
We take advantage of the natural set of units provided by quantum mechanics  for the quantum of action $\hbar=1$, by relativity for the speed of light $c=1$, and by electromagnetism 
 for the vacuum magnetic permeability $\u{\mu_\circ}=1$, 
 \vspace{-.8mm}
 electric permittivity $ \u{\epsilon_\circ}=1$, and vacuum impedance $Z_\circ=$ $(\u{\mu_\circ}/\u{\epsilon_\circ})^{1/2} =1$. 
  \vspace{-.2mm}
We then rescale these natural units, all related to the second or simply equal to 1, to get the properly normalized units we are used to. 

\vspace{2mm}

Deciding a choice of values for the numerical constants $\hbar_\circ,\ c_\circ$, and $\mu_\circ$ (with related $\epsilon_\circ$ and $z_\circ$) {\em does not fix the Planck constant, the speed of light nor the permeability or the permittivity or the impedance of vacuum}, which all remain identical to 1. But it fixes instead the sizes of the various units (with or without dimensions) we have chosen to use, derived from the second. This is also the case for the Boltzmann constant, providing the usual kelvin, and for the Avogadro constant, turned back into a fixed Avogadro number $N_{A\circ}$ providing the definition of the mole.
\vspace{2mm} 

It has been agreed to fix numerically $\hbar_\circ$
and $c_\circ $
to define the metre, the joule, and the kilogram, but decided not to fix $\mu_\circ =4\pi\times 10^{-7}$ (N/A$^2$), to let the coulomb and ampere free to adjust to the chosen value of the elementary charge. This may be practically convenient, but remains conceptually questionable. 
While the vacuum magnetic permeability, electric permittivity and impedance are all equal to 1, as the speed of light, 
their numerical values in SI units (N/A$^2$ or H/m, F/m and $\Omega$, respectively)
get now dependent on the elementary charge $e$,\, i.e. on future experimental measurements of $\alpha$. This is a somewhat unpleasant consequence of the fixing of $e_\circ$ in the new SI. 

\vspace{2mm}

This is well managed by embedding the new SI  within the proposed system, where the floating character of electrical units through their dependence on an unfixed $\mu_\circ$ proportional to $\alpha$  is made explicit and automatically taken care of. $\ \hbar\!=\!c\!=\!\u{\mu_\circ}\!=\!\u{\epsilon_\circ}\!=\!Z_\circ\!=\!k\!=\!N_A\!=\!1$  guarantees that the 
vacuum magnetic permeability, 
electric permittivity and  impedance all remain constant and  identical to 1.
The construction combines  the advantages of both systems, with a simplified description of the fundamental laws, dimensionless 
charges, impedances and fluxes, an elementary charge $\,e=\sqrt{4\pi\alpha}\,$
and flux quantum $\pi/e$, the kelvin as a unit of energy, and the mole identified as a very large Avogadro number.

\end{document}